\documentclass{article}

\usepackage[utf8]{inputenc}
\usepackage[english]{babel}
\usepackage[a4paper, margin=1in]{geometry}
\usepackage{graphicx}
\usepackage{amsmath}
\usepackage{amsfonts}
\usepackage{physics}
\usepackage{gensymb}
\usepackage{bbold}
\usepackage{bm}
\usepackage{cleveref}
\usepackage{tikz}
\usetikzlibrary{matrix}

\newcommand{\hgp}{
    \draw (0,0) -- (4,0) -- (4,4) -- (0,4) -- cycle;
    \draw (2,0) -- (2,4);
    \draw (0,2) -- (4,2);

    \node at (1,1) {$X$};
    \node at (3,3) {$Z$};
    \node at (3,1) {$Q_1$};
    \node at (1,3) {$Q_2$};

    \draw (0,-0.2) -- (4,-0.2) -- (4,-0.4) -- (0,-0.4) -- cycle;
    \draw (2,-0.2) -- (2,-0.4);
    \node at (2,-1) {$H_1$};
    \draw (-0.2,0) -- (-0.2,4) -- (-0.4,4) -- (-0.4,0) -- cycle;
    \draw (-0.2,2) -- (-0.4,2);
    \node [anchor=west] at (-1.5,2) {$H_2$};
    
    \draw [<->] (0.1,-0.6) -- (1.9,-0.6);
    \node at (1,-0.8) {$m_1$};
    \draw [<->] (2.1,-0.6) -- (3.9,-0.6);
    \node at (3,-0.8) {$n_1$};
    \draw [<->] (-0.6,0.1) -- (-0.6,1.9);
    \node [anchor=west] at (-1.3,1) {$n_2$};
    \draw [<->] (-0.6,2.1) -- (-0.6,3.9);
    \node [anchor=west] at (-1.3,3) {$m_2$};
}

\newcommand{\hgpsys}{
    \draw (0,0) -- (4,0) -- (4,4) -- (0,4) -- cycle;
    \draw (2,0) -- (2,4);
    \draw (0,2) -- (4,2);

    \node at (1,1) {$X$};
    \node at (3,3) {$Z$};
    \node at (3,1) {$Q_1$};
    \node at (1,3) {$Q_2$};

    \draw (0,-0.2) -- (4,-0.2) -- (4,-0.4) -- (0,-0.4) -- cycle;
    \draw (2,-0.2) -- (2,-0.4);
    \draw (2.75,-0.2) -- (2.75,-0.4);
    \node at (2,-1) {$H_1$};
    \draw (-0.2,0) -- (-0.2,4) -- (-0.4,4) -- (-0.4,0) -- cycle;
    \draw (-0.2,2) -- (-0.4,2);
    \draw (-0.2,1.25) -- (-0.4,1.25);
    \node [anchor=west] at (-1.5,2) {$H_2$};
   
    \draw (2,1.25) -- (4,1.25);
    \draw (2.75,0) -- (2.75,2);
    \fill [gray, opacity=0.5] (2,1.25) -- (2.75,1.25) -- (2.75,2) -- (2,2) -- cycle;

    \draw [<->] (0.1,-0.6) -- (1.9,-0.6);
    \node at (1,-0.8) {$m_1$};
    \draw [<->] (2.1,-0.6) -- (2.7,-0.6);
    \node at (2.4,-0.8) {$k_1$};
    \draw [<->] (2.8,-0.6) -- (3.9,-0.6);
    \node at (3.4,-0.8) {$n_1-k_1$};
    \draw [<->] (-0.6,0.1) -- (-0.6,1.2);
    \node [anchor=east] at (-0.7,0.8) {$n_2-k_2$};
    \draw [<->] (-0.6,1.3) -- (-0.6,1.9);
    \node [anchor=west] at (-1.3,1.6) {$k_2$};
    \draw [<->] (-0.6,2.1) -- (-0.6,3.9);
    \node [anchor=west] at (-1.3,3) {$m_2$};
}

\newcommand{\lifta}{
    \filldraw (0,0) circle (3pt);
    \filldraw (1,0) circle (3pt);
    \filldraw (2,0) circle (3pt);
    \filldraw (0,1) circle (3pt);
    \filldraw (1,1) circle (3pt);
    \filldraw (2,1) circle (3pt);
    \draw (0,0) -- (1,1) -- (2,0);
    \draw (0,1) -- (1,0) -- (2,1);
}

\newcommand{\liftb}{
    \filldraw (-1,1.5) circle (3pt);
    \filldraw (-1,2.5) circle (3pt);
    \filldraw (-0.5,2.25) circle (3pt);
    \filldraw (-0.5,3.25) circle (3pt);
    \filldraw (0,3) circle (3pt);
    \filldraw (0,4) circle (3pt);
    \draw (-1,1.5) -- (-0.5,3.25) -- (0,3);
    \draw (-1,2.5) -- (-0.5,2.25) -- (0,4);
}

\newcommand{\lp}{
    \draw [gray] (0,0) -- (2,2) -- (2,4);
    \draw [gray] (2,2) -- (6,2);
    \draw [gray] (2,0) -- (4,2);
    \draw [gray] (1,1) -- (5,1);
    \draw [gray] (3,1) -- (3,3);
    \draw [gray] (1,1) -- (1,3);
    \draw [gray] (4,2) -- (4,4);
    \draw (0,2) -- (4,2) -- (6,4) -- (2,4) -- cycle;
    \draw (0,2) -- (0,0) -- (4,0) -- (6,2) -- (6,4);
    \draw (4,0) -- (4,2);
    \draw (2,0) -- (2,2) -- (4,4);
    \draw (1,3) -- (5,3) -- (5,1);

    \node at (1.5,2.5) {$X$};
    \node at (4.5,3.5) {$Z$};
    \node at (3.5,2.5) {$Q_1$};
    \node at (2.5,3.5) {$Q_2$};

    \draw (-0.5,-0.5) rectangle (3.5,1.5);
    \draw (1.5,-0.5) -- (1.5,1.5);
    \draw [<->] (-0.4,-0.7) -- (1.4,-0.7);
    \node at (0.5,-1) {$m_1$};
    \draw [<->] (1.6,-0.-0.7) -- (3.5,-0.7);
    \node at (2.5,-1) {$n_1$};

    \draw (4.5,0) -- (6.5,2) -- (6.5,4) -- (4.5,2) -- cycle;
    \draw (5.5,1) -- (5.5,3);
    \draw [<->] (4.6,-0.1) -- (5.5,0.8);
    \node [anchor=north west] at (5,0.5) {$n_2$};
    \draw [<->] (5.6,0.9) -- (6.5,1.8);
    \node [anchor=north west] at (6,1.5) {$m_2$};

    \draw [<->] (-0.7,-0.5) -- (-0.7,1.5);
    \node at (-1,0.5) {$l$};
}

\newcommand{\lpsimple}{
    \draw [gray] (0,0) -- (2,2) -- (2,4);
    \draw [gray] (2,2) -- (6,2);
    \draw [gray] (2,0) -- (4,2);
    \draw [gray] (1,1) -- (5,1);
    \draw [gray] (3,1) -- (3,3);
    \draw [gray] (1,1) -- (1,3);
    \draw [gray] (4,2) -- (4,4);
    \draw (0,2) -- (4,2) -- (6,4) -- (2,4) -- cycle;
    \draw (0,2) -- (0,0) -- (4,0) -- (6,2) -- (6,4);
    \draw (4,0) -- (4,2);
    \draw (2,0) -- (2,2) -- (4,4);
    \draw (1,3) -- (5,3) -- (5,1);
}

\begin{document}

\title{Visualising Quantum Product Codes}
\author{Tom Scruby}
\date{\small t.r.scruby@gmail.com \\ \textit{Okinawa Institute of Science and Technology}}
\maketitle

\section{Some Things}

\begin{itemize}

\item \textbf{Outline:} Each of the following sections focuses on a type of error correcting code, and each section is split into two parts. The first part simply describes the visualisation I use for this type of code, gives an example, and explains how various properties can be inferred from this visualisation. The second part actually justifies, as rigorously as I am able, why the visualisation is valid. If for some reason you feel inclined to trust what I have written in the first part, feel free to skip the second. 

\item \textbf{Disclaimer:} To the best of my knowledge, these perspectives on lifted and balanced product codes are novel in the sense that they have not previously been publicly described. I do not doubt that they have long been privately known to various members of the community who perhaps have their own notes or perspectives on the topic.

\item \textbf{Etc:} Please let me know if there is anything contained here that is incorrect, unclear, or otherwise in need of amendment. I hope you find these notes useful, and thanks for reading. 

\end{itemize}

\section{Classical Codes}

\subsection{Visualisation}
Just put the vertices of the Tanner graph in a line, checks on the left and bits on the right. We don't learn anything by doing this but we do save a lot of time. Here's an unnecessary example with a three bit repetition code. 

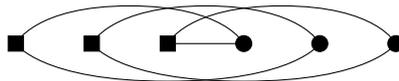
\begin{center}
    \begin{tikzpicture}
        \filldraw (-0.1,-0.1) rectangle (0.1,0.1);
        \filldraw (0.9,-0.1) rectangle (1.1,0.1);
        \filldraw (1.9,-0.1) rectangle (2.1,0.1);
        \filldraw (3,0) circle (3pt);
        \filldraw (4,0) circle (3pt);
        \filldraw (5,0) circle (3pt);

        \draw plot [smooth, tension=1.5] coordinates {(0,0) (1.5,0.5) (3,0)};
        \draw plot [smooth, tension=1.5] coordinates {(1,0) (2.5,0.5) (4,0)};
        \draw plot [smooth, tension=1.5] coordinates {(2,0) (3.5,0.5) (5,0)};
        \draw plot [smooth, tension=1.5] coordinates {(0,0) (2,-0.5) (4,0)};
        \draw plot [smooth, tension=1.5] coordinates {(1,0) (3,-0.5) (5,0)};
        \draw (2,0) -- (3,0);
    \end{tikzpicture}
\end{center}
and in general I will show it like this, where $m$ is the number of checks and $n$ is the number of bits.

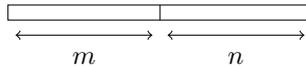
\begin{center}
    \begin{tikzpicture}
        
        \draw (5,0) -- (9,0) -- (9,0.2) -- (5,0.2) -- cycle;
        \draw (7,0) -- (7,0.2);
         
        \draw [<->] (5.1,-0.2) -- (6.9,-0.2);
        \node at (6,-0.5) {$m$};
        \draw [<->] (7.1,-0.2) -- (8.9,-0.2);
        \node at (8,-0.5) {$n$};
    \end{tikzpicture}
\end{center}

It is always possible to find a basis for the codewords of these\footnote{``These'' meaning binary linear codes, which are all we care about for the purposes of these notes.} codes in which, for basis element $x$, all of the first $k$ bits of the code are $0$ except for bit $x$. This gives a correspondence between the first $k$ physical bits and the $k$ logical bits (codes with this property are sometimes called \textit{systematic codes}). Such a basis can be constructed starting from any arbitrary basis via Gaussian elimination. We can include this in our visualisation of the Tanner graph of $H$ like so

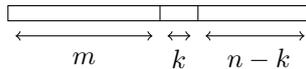
\begin{center}
    \begin{tikzpicture}
        
        \draw (5,0) -- (9,0) -- (9,0.2) -- (5,0.2) -- cycle;
        \draw (7,0) -- (7,0.2);
        \draw (7.5,0) -- (7.5,0.2);
         
        \draw [<->] (5.1,-0.2) -- (6.9,-0.2);
        \node at (6,-0.5) {$m$};
        \draw [<->] (7.1,-0.2) -- (7.4,-0.2);
        \node at (7.25,-0.5) {$k$};
        \draw [<->] (7.6,-0.2) -- (8.9,-0.2);
        \node at (8.3,-0.5) {$n-k$};
    \end{tikzpicture}
\end{center}

\subsection{Justification}
Not really much to say here, mostly just some preliminaries for later. We are specifically concerned with binary linear codes, which are commonly described by parameters $[n,k,d]$ and an $m \times n$ parity check matrix $H$. I will use $i$ to refer to check indices and $j$ to refer to bit indices. The visualisation is achieved by assigning coordinates as follows

\begin{center}
    \begin{tikzpicture}
        \matrix (M) [matrix of nodes, 
                     left delimiter={(}, 
                     right delimiter={)}, 
                     nodes={minimum size=5mm},
                     matrix anchor=west]  at (0,0) 
        {
            \phantom{X} & \phantom{X} & \\
            \phantom{X} & \phantom{X} & \\
        };
        \node [anchor=east] at ([xshift=-3mm]M.west) {$H = $};

        \draw [<->] (M.south west) -- (M.south east);
        \node [anchor=north] at (M.south) {$n$};
        \draw [<->] ([xshift=3mm]M.north east) -- ([xshift=3mm]M.south east);
        \node [anchor=west] at ([xshift=3mm]M.east) {$m$};

        \node at (3.5,0.5) {checks $(i)$};
        \node at (3.5,-0.5) {bits $(j + m)$};
        
        \draw (5,0) -- (9,0) -- (9,0.2) -- (5,0.2) -- cycle;
        \draw (7,0) -- (7,0.2);
         
        \draw [<->] (5.1,-0.2) -- (6.9,-0.2);
        \node at (6,-0.5) {$m$};
        \draw [<->] (7.1,-0.2) -- (8.9,-0.2);
        \node at (8,-0.5) {$n$};

    \end{tikzpicture}
\end{center}

\noindent i.e. we just arrange the vertices of the graph in a line, first checks and then bits, according to their respective row/column index in $H$.

\section{Hypergraph Product Codes}
Original paper arxiv:0903.0566

\subsection{Visualisation}
The visualisation for these codes is very well known, but it's worth reviewing it anyway so we have something to refer back/compare to when visualising fancier products. These codes can be defined graphically (via their Tanner graphs) as the Cartesian product of a pair of Tanner graphs for classical codes. For the product of two copies of the three bit repetition code we get

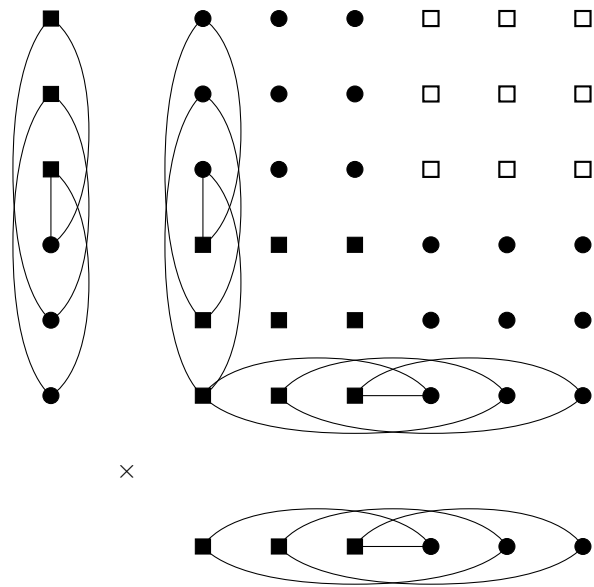
\begin{center}
    \begin{tikzpicture}
        \filldraw (-0.1,-0.1) rectangle (0.1,0.1);
        \filldraw (0.9,-0.1) rectangle (1.1,0.1);
        \filldraw (1.9,-0.1) rectangle (2.1,0.1);
        \filldraw (3,0) circle (3pt);
        \filldraw (4,0) circle (3pt);
        \filldraw (5,0) circle (3pt);

        \draw plot [smooth, tension=1.5] coordinates {(0,0) (1.5,0.5) (3,0)};
        \draw plot [smooth, tension=1.5] coordinates {(1,0) (2.5,0.5) (4,0)};
        \draw plot [smooth, tension=1.5] coordinates {(2,0) (3.5,0.5) (5,0)};
        \draw plot [smooth, tension=1.5] coordinates {(0,0) (2,-0.5) (4,0)};
        \draw plot [smooth, tension=1.5] coordinates {(1,0) (3,-0.5) (5,0)};
        \draw (2,0) -- (3,0);

        \filldraw (-2,2) circle (3pt);
        \filldraw (-2,3) circle (3pt);
        \filldraw (-2,4) circle (3pt);
        \filldraw (-2.1,4.9) rectangle (-1.9,5.1);
        \filldraw (-2.1,5.9) rectangle (-1.9,6.1);
        \filldraw (-2.1,6.9) rectangle (-1.9,7.1);

        \draw plot [smooth, tension=1.5] coordinates {(-2,2) (-1.5,3.5) (-2,5)};
        \draw plot [smooth, tension=1.5] coordinates {(-2,3) (-1.5,4.5) (-2,6)};
        \draw plot [smooth, tension=1.5] coordinates {(-2,4) (-1.5,5.5) (-2,7)};
        \draw plot [smooth, tension=1.5] coordinates {(-2,2) (-2.5,4) (-2,6)};
        \draw plot [smooth, tension=1.5] coordinates {(-2,3) (-2.5,5) (-2,7)};
        \draw (-2,4) -- (-2,5);

        \node at (-1,1) {$\times$};
        
        \begin{scope}[shift={(0,2)}]
            \draw plot [smooth, tension=1.5] coordinates {(0,0) (1.5,0.5) (3,0)};
            \draw plot [smooth, tension=1.5] coordinates {(1,0) (2.5,0.5) (4,0)};
            \draw plot [smooth, tension=1.5] coordinates {(2,0) (3.5,0.5) (5,0)};
            \draw plot [smooth, tension=1.5] coordinates {(0,0) (2,-0.5) (4,0)};
            \draw plot [smooth, tension=1.5] coordinates {(1,0) (3,-0.5) (5,0)};
            \draw (2,0) -- (3,0);
        \end{scope}

        \begin{scope}[shift={(0,2)}]
            \filldraw (-0.1,-0.1) rectangle (0.1,0.1);
            \filldraw (0.9,-0.1) rectangle (1.1,0.1);
            \filldraw (1.9,-0.1) rectangle (2.1,0.1);
            \filldraw (3,0) circle (3pt);
            \filldraw (4,0) circle (3pt);
            \filldraw (5,0) circle (3pt);
        \end{scope}

        \begin{scope}[shift={(0,3)}]
            \filldraw (-0.1,-0.1) rectangle (0.1,0.1);
            \filldraw (0.9,-0.1) rectangle (1.1,0.1);
            \filldraw (1.9,-0.1) rectangle (2.1,0.1);
            \filldraw (3,0) circle (3pt);
            \filldraw (4,0) circle (3pt);
            \filldraw (5,0) circle (3pt);
        \end{scope}

        \begin{scope}[shift={(0,4)}]
            \filldraw (-0.1,-0.1) rectangle (0.1,0.1);
            \filldraw (0.9,-0.1) rectangle (1.1,0.1);
            \filldraw (1.9,-0.1) rectangle (2.1,0.1);
            \filldraw (3,0) circle (3pt);
            \filldraw (4,0) circle (3pt);
            \filldraw (5,0) circle (3pt);
        \end{scope}

        \filldraw (0,5) circle (3pt);
        \filldraw (1,5) circle (3pt);
        \filldraw (2,5) circle (3pt);
        \draw [thick] (2.9,4.9) rectangle (3.1,5.1);
        \draw [thick] (3.9,4.9) rectangle (4.1,5.1);
        \draw [thick] (4.9,4.9) rectangle (5.1,5.1);
        
        \begin{scope}[shift={(2,0)}]
            \draw plot [smooth, tension=1.5] coordinates {(-2,2) (-1.5,3.5) (-2,5)};
            \draw plot [smooth, tension=1.5] coordinates {(-2,3) (-1.5,4.5) (-2,6)};
            \draw plot [smooth, tension=1.5] coordinates {(-2,4) (-1.5,5.5) (-2,7)};
            \draw plot [smooth, tension=1.5] coordinates {(-2,2) (-2.5,4) (-2,6)};
            \draw plot [smooth, tension=1.5] coordinates {(-2,3) (-2.5,5) (-2,7)};
            \draw (-2,4) -- (-2,5);
        \end{scope}

        \begin{scope}[shift={(0,1)}]
            \filldraw (0,5) circle (3pt);
            \filldraw (1,5) circle (3pt);
            \filldraw (2,5) circle (3pt);
            \draw [thick] (2.9,4.9) rectangle (3.1,5.1);
            \draw [thick] (3.9,4.9) rectangle (4.1,5.1);
            \draw [thick] (4.9,4.9) rectangle (5.1,5.1);
        \end{scope}

        \begin{scope}[shift={(0,2)}]
            \filldraw (0,5) circle (3pt);
            \filldraw (1,5) circle (3pt);
            \filldraw (2,5) circle (3pt);
            \draw [thick] (2.9,4.9) rectangle (3.1,5.1);
            \draw [thick] (3.9,4.9) rectangle (4.1,5.1);
            \draw [thick] (4.9,4.9) rectangle (5.1,5.1);
        \end{scope}

    \end{tikzpicture}
\end{center}

Circular nodes (products of two bits or checks) are qubits, filled squares (products of a check and a bit) are $X$ checks and empty squares (products of a bit and a check) are $Z$ checks. The connections along each row/column are inherited from the classical code (only some are shown for visual clarity). I will coarse-grain it like this

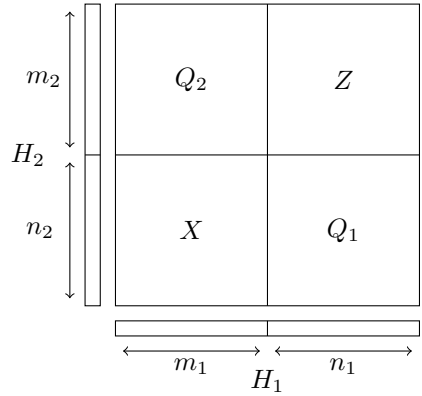
\begin{center}
    \begin{tikzpicture}
       \hgp{} 
    \end{tikzpicture}
\end{center}

Various code parameters and properties can be understood intuitively via this visualisation. For instance:
\newline

\textbf{Stabilisers commute}

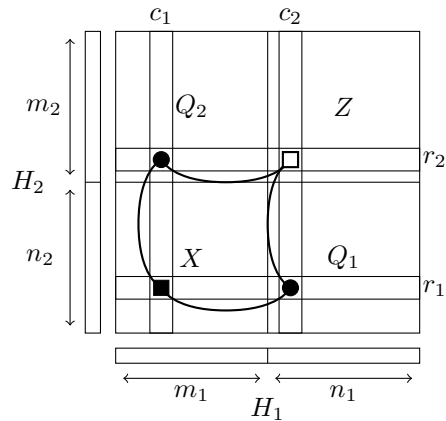
\begin{center}
    \begin{tikzpicture}
        \hgp{}
        \filldraw (2.3,0.6) circle (3pt);
        \filldraw (0.6,2.3) circle (3pt);
        \filldraw (0.5,0.5) rectangle (0.7,0.7);
        \draw [thick] plot [smooth, tension=1.5] coordinates {(0.6,0.6) (1.45,0.3) (2.3,0.6)};
        \draw [thick] plot [smooth, tension=1.5] coordinates {(0.6,0.6) (0.3,1.45) (0.6,2.3)};
        \draw [thick] plot [smooth, tension=1.5] coordinates {(2.3,0.6) (2,1.45) (2.3,2.3)};
        \draw [thick] plot [smooth, tension=1.5] coordinates {(0.6,2.3) (1.45,2) (2.3,2.3)};
        \fill [white] (2.2,2.2) rectangle (2.4,2.4);
        \draw [thick] (2.2,2.2) rectangle (2.4,2.4);
        
        \draw (2.15,0) -- (2.15,4) -- (2.45,4) -- (2.45,0) -- cycle;
        \draw (0,2.15) -- (4,2.15) -- (4,2.45) -- (0,2.45) -- cycle;
        \draw (0.45,0) -- (0.45,4) -- (0.75,4) -- (0.75,0) -- cycle;
        \draw (0,0.45) -- (4,0.45) -- (4,0.75) -- (0,0.75) -- cycle;
        \node at (0.6,4.2) {$c_1$};
        \node at (2.3,4.2) {$c_2$};
        \node at (4.2,0.6) {$r_1$};
        \node at (4.2,2.3) {$r_2$};
    \end{tikzpicture}
\end{center}

This is easy to show. If an $X$ stabiliser with coordiantes $(r_1,c_1)$ and a $Z$ stabiliser with coordinates $(r_2,c_2)$ are both supported on the qubit at $(r_1,c_2)$ in $Q_1$ then they must also both be supported on the qubit at $(r_2,c_1)$ in $Q_2$ because the connectivity in $r_1$ is the same as $r_2$ and similarly for $c_1$ and $c_2$. 
\newline

\textbf{Encoded qubits}

Any codeword of one of the classical codes corresponds to a logical operator of the product, e.g. 

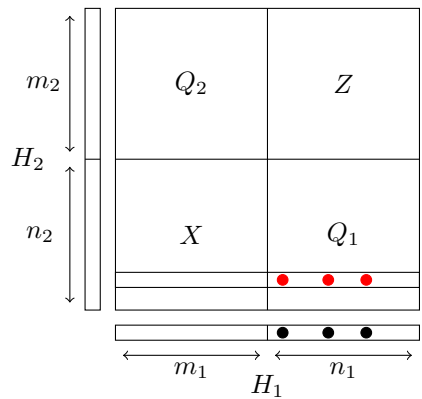
\begin{center}
    \begin{tikzpicture}
        \hgp{}
 
        \draw (0,0.3) -- (4,0.3) -- (4,0.5) -- (0,0.5) -- cycle;
        
        \filldraw [red] (2.2,0.4) circle (2pt);
        \filldraw [red] (2.8,0.4) circle (2pt);
        \filldraw [red] (3.3,0.4) circle (2pt);

        \filldraw (2.2,-0.3) circle (2pt);
        \filldraw (2.8,-0.3) circle (2pt);
        \filldraw (3.3,-0.3) circle (2pt);

    \end{tikzpicture}
\end{center}

Here the black circles are non-zero bits of the classical codeword and red circles are Pauli $Z$ operators on qubits (I will use the convention of red $=Z$, green $=Y$ and blue $=X$ throughout). This operator is guaranteed to commute with all stabilisers because the only $X$ stabilisers that have support overlapping with it correspond to checks of the classical code of which it is a codeword, and all $Z$ stabilisers commute with it because it is a $Z$ operator. Using the fact that the first $k$ bits of the classical code can be made to correspond with the $k$ encoded bits as described in the previous section we can augment our visualisation like so

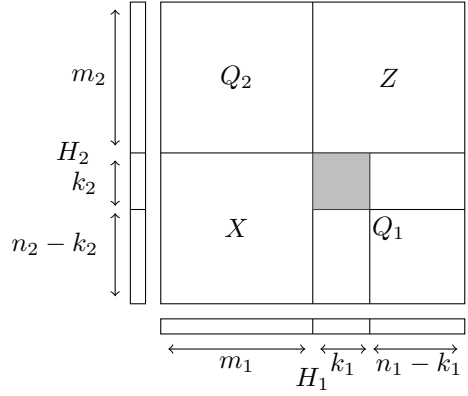
\begin{center}
    \begin{tikzpicture}
        \hgpsys{}        
    \end{tikzpicture}
\end{center}

\noindent where every qubit in the shaded grey area then corresponds to a unique pair of distinct logicals for the quantum code, e.g.

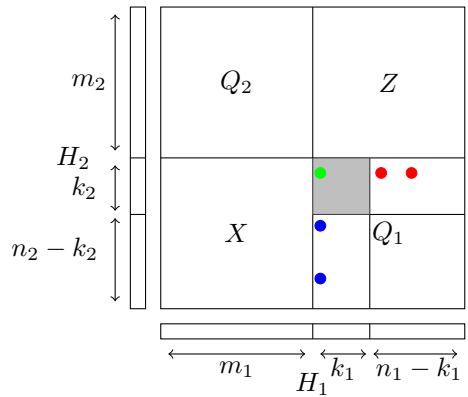
 \begin{center}
    \begin{tikzpicture}
        \hgpsys        
        \filldraw [green] (2.1,1.8) circle (2pt);
        \filldraw [red] (2.9,1.8) circle (2pt);
        \filldraw [red] (3.3,1.8) circle (2pt);
        \filldraw [blue] (2.1,1.1) circle (2pt);
        \filldraw [blue] (2.1,0.4) circle (2pt);

    \end{tikzpicture}
\end{center}

Additionally, we can see that these are all the logicals supported in $Q_1$ because to have more than this we would require the existence of another $X$, $Z$ pair that anticommute with each other but with none of the previous logicals. Such a pair cannot exist because, e.g. for a $Z$ logical like 

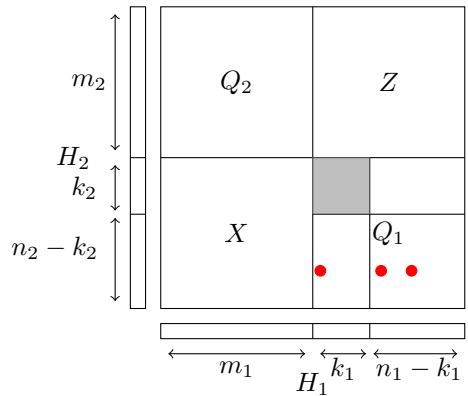
\begin{center}
    \begin{tikzpicture}
        \hgpsys        
        \filldraw [red] (2.1,0.5) circle (2pt);
        \filldraw [red] (2.9,0.5) circle (2pt);
        \filldraw [red] (3.3,0.5) circle (2pt);
    \end{tikzpicture}
\end{center}

\noindent we would need that there are no $X$ logicals associated to a qubit in the grey square with support on the leftmost red qubit, but since all columns are structurally identical this would mean there are no $X$ logicals with support on any qubit of this row, and thus there are no $X$ logicals that anticommute with this operator. This tells us that there are exactly $k_1k_2$ pairs of logicals supported on $Q_1$. 

The structure of the logicals supported on $Q_2$ comes from the structure of the classical codes defined by $H_1^T$ and $H_2^T$ which are just the original codes with bit and check nodes exchanged. In the literature the numbers of logical qubits encoded by these codes are commonly written $k_1^T$ and $k_2^T$, and perfoming the exact same analysis for $Q_2$ we can see that the total number of logical qubits encoded by the hypergraph product code is $k_1k_2 + k_1^Tk_2^T$. 
\newline

\textbf{Code distance}

This is a bit trickier to show, so I will borrow the argument used originally in arxiv:0903.0566. This works by imagining some $Z$ operator that commutes with all $X$ stabilisers but has weight $< \textrm{min}(d_1,d_2^T)$ and support on some subset of columns and rows like so

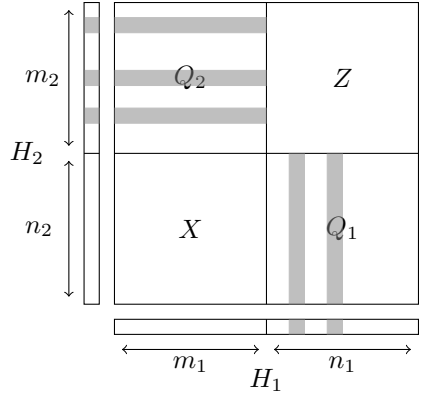
\begin{center}
    \begin{tikzpicture}
        \hgp{}
        \filldraw [gray, opacity=0.5] (2.3,0) rectangle (2.5,2);
        \filldraw [gray, opacity=0.5] (2.3,-0.2) rectangle (2.5,-0.4);
        
        \filldraw [gray, opacity=0.5] (2.8,0) rectangle (3,2);
        \filldraw [gray, opacity=0.5] (2.8,-0.2) rectangle (3,-0.4);

        \filldraw [gray, opacity=0.5] (0,2.4) rectangle (2,2.6);
        \filldraw [gray, opacity=0.5] (-0.2,2.4) rectangle (-0.4,2.6);

        \filldraw [gray, opacity=0.5] (0,2.9) rectangle (2,3.1);
        \filldraw [gray, opacity=0.5] (-0.2,2.9) rectangle (-0.4,3.1);

        \filldraw [gray, opacity=0.5] (0,3.6) rectangle (2,3.8);
        \filldraw [gray, opacity=0.5] (-0.2,3.6) rectangle (-0.4,3.8);
    \end{tikzpicture}
\end{center}

If we consider the punctured classical codes containing all the same checks but only the bits in the grey squares then, because the number of grey columns must be less than $d_1$/number of grey rows must be less than $d_2^T$ these punctured codes have logical dimension zero. Accordingly, the product of these two codes has no logical qubits and so this $Z$ operator must be a stabiliser of this code. Then, because all $Z$ stabilisers of this code are also $Z$ stabilisers of the full code this operator must be a $Z$ stabiliser of the full code as well.
\newline

\textbf{Logical operations}

This visualisation (and in particular the existence of a canonical logical basis) can be used to derive some nice logical operations for these codes. For example, using symmetries for gates by folding (arxiv:2202.06647, 2204.10812) or puncturing classical codes (arxiv:2407.18490). I'm not going to go over all of these here, but the key thing to be aware of is that a nice basis for the logical Pauli operators of the code is necessary for them to have a clearly defined logical action.

\subsection{Justification}
The hypergraph product of a pair of classical codes is commonly defined either as an algebraic operation on a pair of classical parity check matrices or as a visual operation on the corresponding Tanner graphs. Although we are ultimately interested in a purely visual description, in order to generalise this description to other products we first need to clearly establish the connection between the two descriptions and show how the visual one arises from the structure of the algebraic one. We therefore begin by defining the hypergraph product as

\begin{equation}
    \label{eq:hgp}
    H_X = ( H_1 \otimes I_{n_2} ~|~ I_{m_1} \otimes H_2^T ) ~~~~~~ H_Z = ( I_{n_1} \otimes H_2 ~|~ H_1^T \otimes I_{m_2} )
\end{equation}

\noindent where $H_i$ are the $m_i \times n_i$ PCMs of the two classical input codes. It will be helpful to emphasise that, explicitly, for an $m \times n$ matrix $H$ with elements $H[i,j]$ and an $r \times r$ identity matrix $I_r$ we have

\begin{equation}
    \label{eq:matrix_forms}
    H \otimes I_r = \begin{pmatrix}
        H[0,0] I_r & H[0,1] I_r & . & . & . \\
        H[1,0] I_r & H[1,1] I_r & . & . & . \\
        .          & .          &   &   &   \\
        .          & .          &   &   &   \\
        .          & .          &   &   &   \\
    \end{pmatrix}
    ~~~~~~
    I_r \otimes H = \begin{pmatrix}
        H      & \bm{0} & . & . & . \\
        \bm{0} & H      & . & . & . \\
        .      & .      &   &   &   \\
        .      & .      &   &   &   \\
        .      & .      &   &   &   \\
    \end{pmatrix}
\end{equation}
where $\bm{0}$ is the $m \times n$ all-zeros matrix. Elements of these matrices have the forms

\begin{equation}
    \label{eq:element_forms}
        \begin{split}
        &(H \otimes I_r)[i,j] = \begin{cases}
            H[\lfloor i/r \rfloor, \lfloor j/r \rfloor] ~~~ \textrm{ if } i = j \mod r \\
            0 ~~~~~~~~~~~~~~~~~~~~~ \textrm{ otherwise}
        \end{cases}
        \\
        &(I_r \otimes H)[i,j] = \begin{cases}
            H[i \textrm{ mod } m, j \textrm{ mod } n] ~~~ \textrm{ if } \lfloor i/m \rfloor = \lfloor j/n \rfloor \\
            0 ~~~~~~~~~~~~~~~~~~~~~~~~~~~~ \textrm{ otherwise}
        \end{cases}
    \end{split}
\end{equation}

We can use the block structure of these matrices to define an assignment of coordinates to the vertices of the corresponding Tanner graphs. The assignment will depend on the form of the matrix:

\begin{equation}
    \label{eq:2D_coords1}
    \begin{tikzpicture}[baseline=(current bounding box.center)]

        \matrix (M) [matrix of nodes, 
                     left delimiter={(}, 
                     right delimiter={)}, 
                     nodes={minimum size=5mm},
                     matrix anchor=west]  at (-3.5,1.8)
        {
            \phantom{X} & \phantom{X} & \phantom{X} \\
            \phantom{X} & \phantom{X} & \phantom{X} \\
            \phantom{X} & \phantom{X} & \phantom{X} \\
        };
        \node [anchor=east] at ([xshift=-3mm]M.west) {$H \otimes I_r = $};
  
        \draw (M-1-1.south west) -- (M-1-3.south east);
        \draw (M-1-1.north east) -- (M-3-1.south east);
        \draw [<->] (M-2-1.west) -- (M-2-1.east);
        \node at (M-2-1.south) {$r$};
        \draw [<->] (M-1-2.north) -- (M-1-2.south);
        \node at (M-1-2.east) {$r$};
        \draw [<->] (M.south west) -- (M.south east);
        \node [anchor=north] at (M.south) {$nr$};
        \draw [<->] ([xshift=3mm]M.north east) -- ([xshift=3mm]M.south east);
        \node [anchor=west] at ([xshift=3mm]M.east) {$mr$};

        \node at (-3.5,0.2) {$\textrm{checks } (\lfloor i/r \rfloor, i \textrm{ mod } r)$};
        \node at (-3.5,-0.5) {$\textrm{bits } (\lfloor j/r \rfloor + m, j \textrm{ mod } r)$};

        \draw (0,0) -- (4,0) -- (4,2) -- (0,2) -- cycle;
        \draw (2,0) -- (2,2);
        \draw [<->] (0.1,-0.2) -- (1.9,-0.2);
        \node at (1,-0.5) {$m$};
        \draw [<->] (2.1,-0.2) -- (3.9,-0.2);
        \node at (3,-0.5) {$n$};
        \draw [<->] (4.2,0.1) -- (4.2,1.9);
        \node at (4.5,1) {$r$};

        \filldraw [gray, opacity=0.5] (0,0) -- (4,0) -- (4,0.2) -- (0,0.2) -- cycle;

    \end{tikzpicture}
\end{equation}
   
\begin{equation}
    \label{eq:2D_coords2}
    \begin{tikzpicture}[baseline=(current bounding box.center)]
        
        \matrix (M) [matrix of nodes, 
                     left delimiter={(}, 
                     right delimiter={)}, 
                     nodes={minimum size=5mm},
                     matrix anchor=west]  at (-3.5,3.2)
        {
            \phantom{X} & \phantom{X} & \phantom{X} \\
            \phantom{X} & \phantom{X} & \phantom{X} \\
            \phantom{X} & \phantom{X} & \phantom{X} \\
        };
        \node [anchor=east] at ([xshift=-3mm]M.west) {$I_r \otimes H = $};
  
        \draw (M-1-1.south west) -- (M-1-3.south east);
        \draw (M-1-1.north east) -- (M-3-1.south east);
        \draw [<->] (M-2-1.west) -- (M-2-1.east);
        \node at (M-2-1.south) {$n$};
        \draw [<->] (M-1-2.north) -- (M-1-2.south);
        \node at (M-1-2.east) {$m$};
        \draw [<->] (M.south west) -- (M.south east);
        \node [anchor=north] at (M.south) {$rn$};
        \draw [<->] ([xshift=3mm]M.north east) -- ([xshift=3mm]M.south east);
        \node [anchor=west] at ([xshift=3mm]M.east) {$rm$};

        \node at (-3.5,1.3) {$\textrm{checks } (\lfloor i/m \rfloor, i \textrm{ mod } m)$};
        \node at (-3.5,0.3) {$\textrm{bits } (\lfloor j/n \rfloor, (j \textrm{ mod } n) + m)$};

        \draw (0,0) -- (2,0) -- (2,4) -- (0,4) -- cycle;
        \draw (0,2) -- (2,2);
        \draw [<->] (0.1,-0.2) -- (1.9, -0.2);
        \node at (1,-0.5) {$r$};
        \draw [<->] (2.2,0.1) -- (2.2,1.9);
        \node at (2.5,1) {$m$};
        \draw [<->] (2.2,2.1) -- (2.2,3.9);
        \node at (2.5,3) {$n$};
        
        \filldraw [gray, opacity=0.5] (0,0) -- (0,4) -- (0.2,4) -- (0.2,0) -- cycle;

    \end{tikzpicture}
\end{equation}

A useful property of this visualisation is that rows of vertices with a common $y$ coordinate/columns of vertices with a common $r$ coordinate look like copies of the Tanner graph of $H$ (grey rectangles in the above images). We can convince ourselves of this by first observing that, by \cref{eq:element_forms}, elements of $H \otimes I_r$ can be non-zero only when $i = j \mod r$, meaning (from \cref{eq:2D_coords1}) vertices in the visualisation can only be connected if they share a common $y$ coordinate. If we fix such a coordinate then the solutions of 
\begin{equation}
    y = i = j \mod r ~~~ y \in [0,r)
\end{equation}
have the form 
\begin{equation}
    \begin{split}
        &i = y + ar ~~~ a \in [0,m) \\
        &j = y + br ~~~ b \in [0,n)
    \end{split}
\end{equation}
Then, again from \cref{eq:element_forms} we see that each set of solutions gives us a set of elements 
\begin{equation}
    H[\lfloor (y + ar)/r \rfloor, \lfloor (y + br)/r \rfloor] = H[a,b] 
\end{equation}
so we have one copy of $H$ for each choice of $y$, giving one copy of the Tanner graph of $H$ for each row in the visualisation of $H \otimes I_r$. Similarly, for $I_r \otimes H$ elements can only be non-zero if $\lfloor i/m \rfloor = \lfloor j/n \rfloor$ and from \cref{eq:2D_coords2} this means connected vertices in the visualisation must share a common $x$ coordinate. For any value of this coordinate the solutions of 
\begin{equation}
    x = \lfloor i/m \rfloor = \lfloor j/n \rfloor ~~~ x \in [0,r)
\end{equation}
have the form
\begin{equation}
    \begin{split}
        i = a + xm ~~~ a \in [0,m) \\
        j = b + xn ~~~ b \in [0,n)
    \end{split}
\end{equation}
\Cref{eq:element_forms} then tells us each set of solutions corresponds to a set of elements
\begin{equation}
    H[(a + xm) \textrm{ mod } m, (b + xn) \textrm{ mod } n] = H[a,b]
\end{equation}
so we have one copy of $H$ for each choice of $x$ and one copy of the Tanner graph of $H$ for each column in the visualisation of $H \otimes I_r$. 

We can now apply this visualisation to $H_X$ and $H_Z$ as given in \cref{eq:hgp}. Focusing first on the two halves of $H_X$ we have a visualisation of the $X$ Tanner graph of the code with the following layout (remembering that for $I_{m_1} \otimes H_2^T$ we have $m$ and $n$ exchanged relative to \cref{eq:2D_coords2} because $H_2^T$ is transpose)

\begin{center}
    \begin{tikzpicture}

        \matrix (M1) [matrix of nodes, 
                     left delimiter={(}, 
                     right delimiter={)}, 
                     nodes={minimum size=5mm},
                     matrix anchor=west]  at (-4,3.1)
        {
            \phantom{X} & \phantom{X} & \phantom{X} \\
            \phantom{X} & \phantom{X} & \phantom{X} \\
            \phantom{X} & \phantom{X} & \phantom{X} \\
        };
        \node [anchor=east] at ([xshift=-3mm]M1.west) {$H_1 \otimes I_{n_2} = $};

        \draw (M1-1-1.south west) -- (M1-1-3.south east);
        \draw (M1-1-1.north east) -- (M1-3-1.south east);
        \draw [<->] (M1-2-1.west) -- (M1-2-1.east);
        \node at (M1-2-1.south) {$n_2$};
        \draw [<->] (M1-1-2.north) -- (M1-1-2.south);
        \node at (M1-1-2.east) {$n_2$};
        \draw [<->] (M1.south west) -- (M1.south east);
        \node [anchor=north] at (M1.south) {$n_1n_2$};
        \draw [<->] ([xshift=3mm]M1.north east) -- ([xshift=3mm]M1.south east);
        \node [anchor=west] at ([xshift=3mm]M1.east) {$m_1n_2$};

        \matrix (M2) [matrix of nodes, 
                     left delimiter={(}, 
                     right delimiter={)}, 
                     nodes={minimum size=5mm},
                     matrix anchor=west]  at (-4,0.7)
        {
            \phantom{X} & \phantom{X} & \phantom{X} \\
            \phantom{X} & \phantom{X} & \phantom{X} \\
            \phantom{X} & \phantom{X} & \phantom{X} \\
        };
        \node [anchor=east] at ([xshift=-3mm]M2.west) {$I_{m_1} \otimes H_2^T = $};

        \draw (M2-1-1.south west) -- (M2-1-3.south east);
        \draw (M2-1-1.north east) -- (M2-3-1.south east);
        \draw [<->] (M2-2-1.west) -- (M2-2-1.east);
        \node at (M2-2-1.south) {$m_2$};
        \draw [<->] (M2-1-2.north) -- (M2-1-2.south);
        \node at (M2-1-2.east) {$n_2$};
        \draw [<->] (M2.south west) -- (M2.south east);
        \node [anchor=north] at (M2.south) {$m_1m_2$};
        \draw [<->] ([xshift=3mm]M2.north east) -- ([xshift=3mm]M2.south east);
        \node [anchor=west] at ([xshift=3mm]M2.east) {$m_1n_2$};

        \draw (0,0) -- (4,0) -- (4,2) -- (0,2) -- cycle;
        \draw (0,0) -- (0,4) -- (2,4) -- (2,0) -- cycle;
        \draw [<->] (0,-0.2) -- (2,-0.2);
        \node at (1,-0.5) {$m_1$};
        \draw [<->] (2,-0.2) -- (4,-0.2);
        \node at (3,-0.5) {$n_1$};
        \draw [<->] (-0.2,0) -- (-0.2,2);
        \node at (-0.5,1) {$n_2$};
        \draw [<->] (-0.2,2) -- (-0.2,4);
        \node at (-0.5,3) {$m_2$};

        \node at (1,1) {$X$};
        \node at (3,1) {$Q_1$};
        \node at (1,3) {$Q_2$};

        \node [anchor=west] at (4.2,3) {$X:$ $(\lfloor i/n_2 \rfloor, i \textrm{ mod } n_2)$};
        \node [anchor=west] at (4.2,2) {$Q_1:$ $(\lfloor j/n_2 \rfloor + m_1, j \textrm{ mod } n_2)$};
        \node [anchor=west] at (4.2,1) {$Q_2:$ $(\lfloor j/m_2 \rfloor, (j \textrm{ mod } m_2) + n_2)$};

    \end{tikzpicture}
\end{center}

Here $X$ contains all the vertices of the Tanner graph corresponding to the $X$ checks while $Q_1$ are the qubit vertices corresponding to columns on the left half of $H_X$ and $Q_2$ are those corresponding to columns on the right. The visualisations of $H_1 \otimes I_{n_2}$ and $I_{m_1} \otimes H_2^T$ can be overlayed because the check coordinates cooincide in both cases. 

We can then do the same thing with $H_Z$ 

\begin{center}
    \begin{tikzpicture}
        \draw (0,0) -- (4,0) -- (4,2) -- (0,2) -- cycle;
        \draw (0,0) -- (0,4) -- (2,4) -- (2,0) -- cycle;
        \draw [<->] (0,-0.2) -- (2,-0.2);
        \node at (1,-0.5) {$n_1$};
        \draw [<->] (2,-0.2) -- (4,-0.2);
        \node at (3,-0.5) {$m_1$};
        \draw [<->] (-0.2,0) -- (-0.2,2);
        \node at (-0.5,1) {$m_2$};
        \draw [<->] (-0.2,2) -- (-0.2,4);
        \node at (-0.5,3) {$n_2$};

        \node at (1,1) {$Z$};
        \node at (1,3) {$Q_1$};
        \node at (3,1) {$Q_2$};

        \node [anchor=west] at (5,3) {$Z:$ $(\lfloor i/m_2 \rfloor, i \textrm{ mod } m_2)$};
        \node [anchor=west] at (5,2) {$Q_1:$ $(\lfloor j/n_2 \rfloor, (j \textrm{ mod } n_2) + m_2)$};
        \node [anchor=west] at (5,1) {$Q_2:$ $(\lfloor j/m_2 \rfloor + n_1, j \textrm{ mod } m_2$};

    \end{tikzpicture}
\end{center}

However, the qubit coordinates qubits do not coincide between the visualations of $H_X$ and $H_Z$, whereas the check coordinates do. We would like the opposite to be true so we modify the coordinate assignment to $H_Z$ in the following way 

\begin{center}
    \begin{tikzpicture}
        \matrix (M) [matrix of nodes, 
                     left delimiter={(}, 
                     right delimiter={)}, 
                     nodes={minimum size=5mm},
                     matrix anchor=west]  at (0,1)
        {
            \phantom{X} & \phantom{X} & \phantom{X} \\
            \phantom{X} & \phantom{X} & \phantom{X} \\
            \phantom{X} & \phantom{X} & \phantom{X} \\
        };
        \node [anchor=east] at ([xshift=-3mm]M.west) {$I_{n_1} \otimes H_2 = $};
  
        \draw (M-1-1.south west) -- (M-1-3.south east);
        \draw (M-1-1.north east) -- (M-3-1.south east);
        \draw [<->] (M-2-1.west) -- (M-2-1.east);
        \node at (M-2-1.south) {$n_2$};
        \draw [<->] (M-1-2.north) -- (M-1-2.south);
        \node at (M-1-2.east) {$m_2$};
        \draw [<->] (M.south west) -- (M.south east);
        \node [anchor=north] at (M.south) {$n_1n_2$};
        \draw [<->] ([xshift=3mm]M.north east) -- ([xshift=3mm]M.south east);
        \node [anchor=west] at ([xshift=3mm]M.east) {$n_1m_2$};

        \node [anchor=west] at (3.5,1.5) {checks $(\lfloor i/m_2 \rfloor, i \textrm{ mod } m_2) ~ \rightarrow ~ (\lfloor i/m_2 \rfloor + m_1, (i \textrm{ mod } m_2) + n_2)$};
        \node [anchor=west] at (3.5,0.5) {bits $(\lfloor j/n_2 \rfloor, (j \textrm{ mod } n_2) + m_2) ~ \rightarrow ~ (\lfloor j/n_2 \rfloor + m_1, j \textrm{ mod } n_2)$};

    \end{tikzpicture}
\end{center}

\begin{center}
    \begin{tikzpicture}
        \matrix (M) [matrix of nodes, 
                     left delimiter={(}, 
                     right delimiter={)}, 
                     nodes={minimum size=5mm},
                     matrix anchor=west]  at (0,1)
        {
            \phantom{X} & \phantom{X} & \phantom{X} \\
            \phantom{X} & \phantom{X} & \phantom{X} \\
            \phantom{X} & \phantom{X} & \phantom{X} \\
        };
        \node [anchor=east] at ([xshift=-3mm]M.west) {$H_1^T \otimes I_{m_2} = $};
  
        \draw (M-1-1.south west) -- (M-1-3.south east);
        \draw (M-1-1.north east) -- (M-3-1.south east);
        \draw [<->] (M-2-1.west) -- (M-2-1.east);
        \node at (M-2-1.south) {$m_2$};
        \draw [<->] (M-1-2.north) -- (M-1-2.south);
        \node at (M-1-2.east) {$m_2$};
        \draw [<->] (M.south west) -- (M.south east);
        \node [anchor=north] at (M.south) {$m_1m_2$};
        \draw [<->] ([xshift=3mm]M.north east) -- ([xshift=3mm]M.south east);
        \node [anchor=west] at ([xshift=3mm]M.east) {$n_1m_2$};

        \node [anchor=west] at (3.5,1.5) {checks $(\lfloor i/m_2 \rfloor, i \textrm{ mod } m_2) ~ \rightarrow ~ (\lfloor i/m_2 \rfloor + m_1, (i \textrm{ mod } m_2) + n_2)$}; 
        \node [anchor=west] at (3.5,0.5) {bits $(\lfloor j/r2 \rfloor + n_1, j \textrm{ mod } m_2) ~ \rightarrow ~ (\lfloor j/m_2 \rfloor, (j \textrm{ mod } m_2) + n_2)$};

    \end{tikzpicture}
\end{center}

\begin{center}
    \begin{tikzpicture}
        \draw (2,0) -- (2,4) -- (4,4) -- (4,0) -- cycle;
        \draw (0,2) -- (4,2) -- (4,4) -- (0,4) -- cycle;
        \draw [<->] (0,-0.2) -- (2,-0.2);
        \node at (1,-0.5) {$m_1$};
        \draw [<->] (2,-0.2) -- (4,-0.2);
        \node at (3,-0.5) {$n_1$};
        \draw [<->] (-0.2,0) -- (-0.2,2);
        \node at (-0.5,1) {$n_2$};
        \draw [<->] (-0.2,2) -- (-0.2,4);
        \node at (-0.5,3) {$m_2$};

        \node at (3,3) {$Z$};
        \node at (3,1) {$Q_1$};
        \node at (1,3) {$Q_2$};

        \node [anchor=west] at (5,3) {$Z:$ $(\lfloor i/m_2 \rfloor + m_1, (i \textrm{ mod } m_2) + n_2)$};
        \node [anchor=west] at (5,2) {$Q_1:$ $(\lfloor j/n_2 \rfloor + m_1, j \textrm{ mod } n_2)$};
        \node [anchor=west] at (5,1) {$Q_2:$ $(\lfloor j/m_2 \rfloor, (j \textrm{ mod } m_2) + n_2)$};

    \end{tikzpicture}
\end{center}

\noindent i.e. we exchange the bit and check blocks and then displace the whole thing by $(m_1,0)$ or $(0,n_2)$ respectively. We can see that this preserves the coincidence of the check coordinates and that the qubit coordinates now coincide with those from the $H_X$ visualisation. Altogether we have 

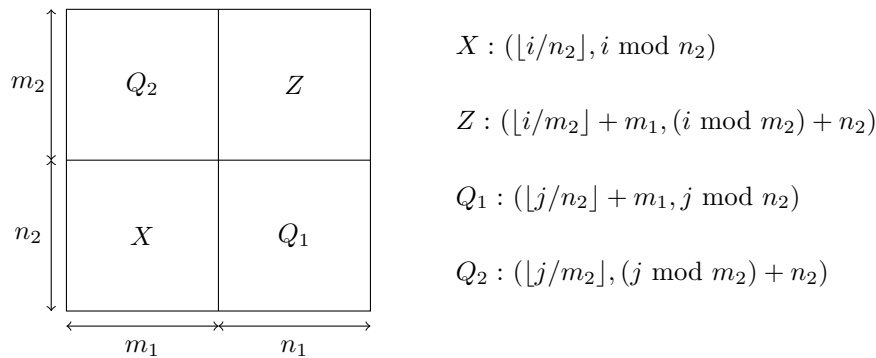
\begin{center}
    \begin{tikzpicture}
        \draw (0,0) -- (4,0) -- (4,4) -- (0,4) -- cycle;
        \draw (2,0) -- (2,4);
        \draw (0,2) -- (4,2);
        \draw [<->] (0,-0.2) -- (2,-0.2);
        \node at (1,-0.5) {$m_1$};
        \draw [<->] (2,-0.2) -- (4,-0.2);
        \node at (3,-0.5) {$n_1$};
        \draw [<->] (-0.2,0) -- (-0.2,2);
        \node at (-0.5,1) {$n_2$};
        \draw [<->] (-0.2,2) -- (-0.2,4);
        \node at (-0.5,3) {$m_2$};

        \node at (1,1) {$X$};
        \node at (3,3) {$Z$};
        \node at (3,1) {$Q_1$};
        \node at (1,3) {$Q_2$};

        \node [anchor=west] at (5,3.5) {$X:$ $(\lfloor i/n_2 \rfloor, i \textrm{ mod } n_2)$};
        \node [anchor=west] at (5,2.5) {$Z:$ $(\lfloor i/m_2 \rfloor + m_1, (i \textrm{ mod } m_2) + n_2)$};
        \node [anchor=west] at (5,1.5) {$Q_1:$ $(\lfloor j/n_2 \rfloor + m_1, j \textrm{ mod } n_2)$};
        \node [anchor=west] at (5,0.5) {$Q_2:$ $(\lfloor j/m_2 \rfloor, (j \textrm{ mod } m_2) + n_2)$};

    \end{tikzpicture}
\end{center}

\section{Lifted Product Codes}
Original paper arxiv:2012.04068

\subsection{Visualisation}
A lift (or covering graph) of a graph $G$ is another graph, $C$, with a map $f$ from the vertices of $C$ to those of $G$ such that the neighbourhood of a vertex $v$ of $C$ is mapped bijectively into the neighbourhood of $f(v)$ in $G$. An $l$-lift is a lift where exactly $l$ vertices of $C$ are mapped to each vertex of $G$. Here is an example with a line graph and a $2$-lift:

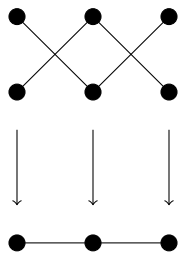
\begin{center}
    \begin{tikzpicture}
        \filldraw (0,0) circle (3pt);
        \filldraw (1,0) circle (3pt);
        \filldraw (2,0) circle (3pt);
        \draw (0,0) -- (2,0);
        
        \begin{scope}[shift={(0,2)}]
        \lifta{}
        \end{scope}
 
        \draw [->] (0,1.5) -- (0,0.5);
        \draw [->] (1,1.5) -- (1,0.5);
        \draw [->] (2,1.5) -- (2,0.5);
    \end{tikzpicture}
\end{center}

\noindent so the top two vertices in each column are mapped to the vertex at the bottom. 

The lifted product makes use of techniques to generate lifts of a chosen graph, called the base graph. These will be explained properly in the next section but for now its enough to know that they exist. They will be visualised as above, i.e. they will be drawn in 2D so that the vertices in each column are mapped to the same vertex of the base graph by the covering map. The lifted product can then be thought of as a kind of 3D Cartesian product of these two 2D graphs (this probably has a proper name in the graph theory literature but I don't know what it is), e.g.

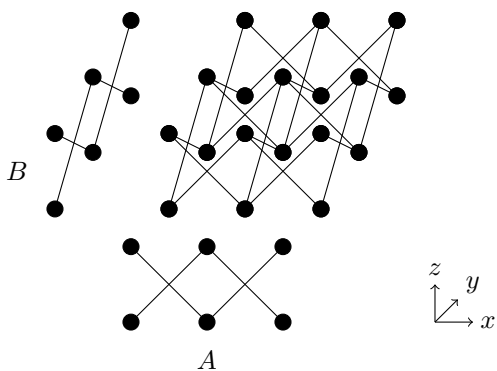
\begin{center}
    \begin{tikzpicture}
        \lifta
        \node at (1,-0.5) {$A$};

        \liftb
        \node at (-1.5,2) {$B$};

        \begin{scope}[shift={(0.5,1.5)}]
            \lifta{}
        \end{scope}
        \begin{scope}[shift={(1,2.25)}]
            \lifta{}
        \end{scope}
        \begin{scope}[shift={(1.5,3)}]
            \lifta{}
        \end{scope}

        \begin{scope}[shift={(1.5,0)}]
            \liftb{}
        \end{scope}
        \begin{scope}[shift={(2.5,0)}]
            \liftb{}
        \end{scope}
        \begin{scope}[shift={(3.5,0)}]
            \liftb{}
        \end{scope}

        \draw [->] (4,0) -- (4.5,0);
        \draw [->] (4,0) -- (4.3,0.3);
        \draw [->] (4,0) -- (4,0.5);
        \node at (4.7,0) {$x$};
        \node at (4.5,0.5) {$y$};
        \node at (4,0.7) {$z$};
 
    \end{tikzpicture}
\end{center}

\noindent so every $xz$-plane of vertices looks like a copy of $A$ and every $yz$-plane of vertices looks like a copy of $B$. Note that, as far as I'm aware, all the work on lifted product codes in the literature uses $l$-lifts and I'm not sure if it is possible to use more general lifts within the current mathematical framework. 

When the base graph is a Tanner graph of a classical code then we have the additional requirement on the lift that the covering map only maps check vertices of the lift to check vertices of the base graph and similarly for bits. Using these lifts as inputs to the 3D product we then assign vertices of the output to be either qubits (product of two bit or two check vertices), $X$ checks (product of a check vertex from the first graph and a bit vertex from the second) or $Z$ checks (product of a bit vertex from the first graph and a check vertex from the second) just as in the HGP.

General lifted product codes can then be represented like this:

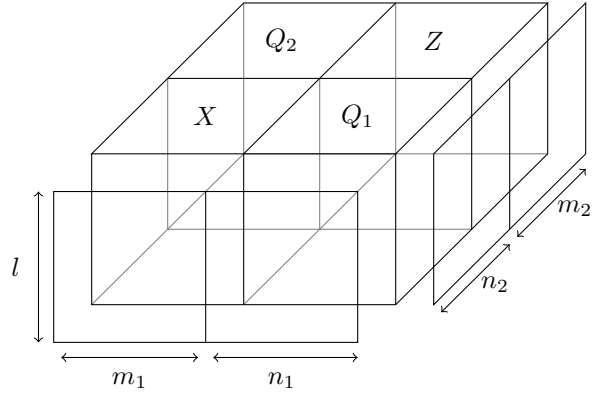
\begin{center}
    \begin{tikzpicture}
        \lp{}
    \end{tikzpicture}
\end{center}

\noindent where $m_1$ and $n_1$ ($m_2$ and $n_2$) are the numbers of checks and bits in the first (second) base graph input to the product and $l$ is the size of the lift (which must be the same for the two graphs). 

Notice that we can perfom the lift and the product in either order, i.e. we can either perform an $l$-lift of the two base graphs and then take the 3D product of these, or we can take the regular Cartesian product of the two base graphs and then perform an $l$-lift on the result. This latter perspective is similar to the perspective on (a subset of) lifted product codes presented in arxiv:2401.02911. It may also be related to arxiv:2404.16736 but I don't understand that paper well enough to say. 
\newline

\textbf{Savings relative to HGP}

The lifted classical codes are obviously still just graphs and so rather than stacking them up in 2D we could stretch them out in a line and take the regular Cartesian/hypergraph product. The two classical codes would then have $m_1l$ and $m_2l$ checks and $n_1l$ and $n_2l$ bits, and so the product would have total dimensions $(m_1 + n_1)l \times (m_2 + n_2)l = (m_1 + n_1)(m_2 + n_2)l^2$. By contrast, the lifted product has total dimensions $(m_1 + n_1)(m_2 + n_2)l$ and so we save a factor of $l$ relative to the hypergraph product. 
\newline

\textbf{Checks do not necessarily commute}

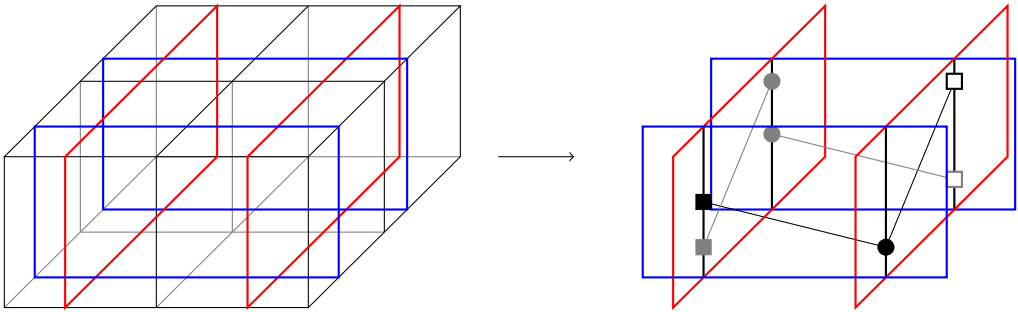
\begin{center}
    \begin{tikzpicture}
        \lpsimple{}
        
        \draw [red, thick] (3.2,0) -- (5.2,2) -- (5.2,4) -- (3.2,2) -- cycle;
        \draw [blue, thick] (1.3,1.3) -- (5.3,1.3) -- (5.3,3.3) -- (1.3,3.3) -- cycle;
        \draw [red, thick] (0.8, 0) -- (2.8,2) -- (2.8,4) -- (0.8,2) -- cycle;
        \draw [blue, thick] (0.4,0.4) -- (4.4,0.4) -- (4.4,2.4) -- (0.4,2.4) -- cycle;

        \draw [->] (6.5,2) -- (7.5,2);
        
        \begin{scope}[shift = {(8,0)}]

            \draw [blue, thick] (1.3,1.3) -- (5.3,1.3) -- (5.3,3.3) -- (1.3,3.3) -- cycle;

            \draw (1.2,1.4) -- (3.6,0.8);
            \draw (4.5,3) -- (3.6,0.8);
            
            \draw [thick] (1.2,0.4) -- (1.2,2.4);
            \draw [thick] (2.1,1.3) -- (2.1,3.3);
            \draw [thick] (4.5,1.3) -- (4.5,3.3);

            \draw [red, thick] (0.8, 0) -- (2.8,2) -- (2.8,4) -- (0.8,2) -- cycle;

            \filldraw (1.1,1.3) rectangle (1.3,1.5);
            \filldraw (3.6,0.8) circle (3pt);
            \fill [white] (4.4,2.9) rectangle (4.6,3.1);    
            \draw [thick] (4.4,2.9) rectangle (4.6,3.1);    

            \filldraw [gray] (2.1,3) circle (3pt);
            \filldraw [gray] (1.1,0.7) rectangle (1.3,0.9);
            \draw [gray] (2.1,3) -- (1.2,0.8);

            \draw [gray] (2.1,2.3) -- (4.5,1.7);

            \filldraw [gray] (2.1,2.3) circle (3pt);
            \fill [white] (4.4,1.6) rectangle (4.6,1.8);
            \draw [thick, gray] (4.4,1.6) rectangle (4.6,1.8);
                    
            \draw [thick] (3.6,0.4) -- (3.6,2.4);

            \draw [red, thick] (3.2,0) -- (5.2,2) -- (5.2,4) -- (3.2,2) -- cycle;
            \draw [blue, thick] (0.4,0.4) -- (4.4,0.4) -- (4.4,2.4) -- (0.4,2.4) -- cycle;
        \end{scope}

    \end{tikzpicture}
\end{center}

On the left we choose four planes through the lifted product code, analagous to the two rows and columns we chose through the hypergraph product code when thinking about commutivity of stabilisers in that case. On the right we restrict to just these planes (for visual clarity) and choose a pair of $X$ and $Z$ checks (black solid square and black empty square) which intersect at a qubit in $Q_1$ (black circle). Because the connectivities in the two blue planes are identical, and so are the connectivities in the two red planes, we must have the grey qubits and checks as well. However, unlike in the HGP case this does not guarantee any extra intersection of the original $X$ and $Z$ checks on qubits in $Q_2$. Additional structure is therefore required to achieve commutivity of $X$ and $Z$ checks in these codes (Pantaleev and Kalachev use ``elementwise commuting matrices'', for example).
\newline

\textbf{Encoded qubits}

The approach to calculating the number of encoded qubits in HGP codes does not generalise to this setting. As with the HGP codes, any codeword of one of the classical (lifted) inputs corresponds to an operator that commutes with all checks of the quantum code. However, it is not so easy to count the number of distinct logical operator pairs. For instance, imagine we could put the lifted classical code into an analogy of the systematic form used for the HGP inputs, which might look something like

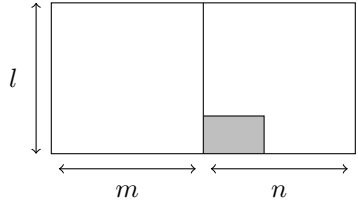
\begin{center}
    \begin{tikzpicture}
        \fill [gray, opacity=0.5] (2,0) rectangle (2.8,0.5);
        \draw (0,0) rectangle (4,2);
        \draw (2,0) -- (2,2);
        \draw (2,0.5) -- (2.8,0.5) -- (2.8,0);

        \draw [<->] (0.1,-0.2) -- (1.9,-0.2);
        \draw [<->] (2.1,-0.2) -- (3.9,-0.2);
        \draw [<->] (-0.2,0) -- (-0.2,2);

        \node at (1,-0.5) {$m$};
        \node at (3,-0.5) {$n$};
        \node at (-0.5,1) {$l$};
    \end{tikzpicture}
\end{center}

\noindent so that each bit in the shaded area corresponds to one of the $k$ logical bits of the code. We might then expect the product to have a structure like

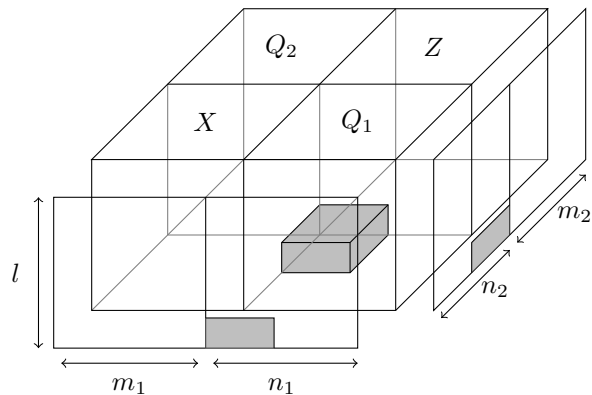
\begin{center}
    \begin{tikzpicture}

        \fill [gray, opacity=0.5] (2.5,0.5) rectangle (3.4,0.9);
        \fill [gray, opacity=0.5] (3.4,0.5) -- (3.9,1) -- (3.9,1.4) -- (3.4,0.9) -- cycle;
        \fill [gray, opacity=0.5] (2.5,0.9) -- (3.4,0.9) -- (3.9,1.4) -- (3,1.4) -- cycle;

        \draw (2.5,0.5) -- (3.4,0.5) -- (3.9,1) -- (3.9,1.4) -- (3,1.4) -- (2.5,0.9) -- cycle;
        \draw (2.5,0.9) -- (3.4,0.9) -- (3.9,1.4);
        \draw (3.4,0.5) -- (3.4,0.9);

        \lp{}

        \fill [gray, opacity=0.5] (1.5,-0.5) rectangle (2.4,-0.1);
        \draw (1.5,-0.1) -- (2.4,-0.1) -- (2.4,-0.5);

        \fill [gray, opacity=0.5] (5,0.5) -- (5.5,1) -- (5.5,1.4) -- (5,0.9) -- cycle;
        \draw (5.5,1.4) -- (5,0.9) -- (5,0.5);

    \end{tikzpicture}
\end{center}

\noindent where each qubit in the shaded volume corresponds to a logical qubit of the code. Unfortunately this is not necessarily true. To see why, consider the following example 

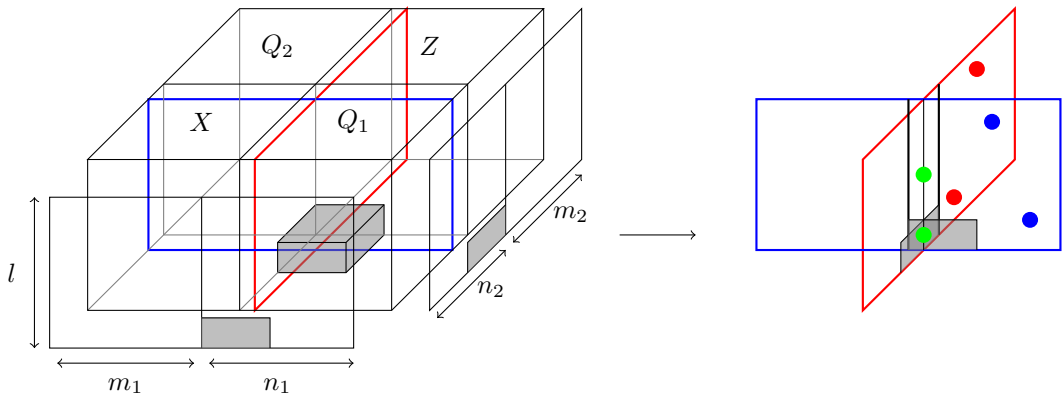
\begin{center}
    \begin{tikzpicture}
        
        \draw [red,thick] (2.2,0) -- (4.2,2) -- (4.2,4) -- (2.2,2) -- cycle;
        \draw [blue,thick] (0.8,0.8) -- (4.8,0.8) -- (4.8,2.8) -- (0.8,2.8) -- cycle;

        \fill [gray, opacity=0.5] (2.5,0.5) rectangle (3.4,0.9);
        \fill [gray, opacity=0.5] (3.4,0.5) -- (3.9,1) -- (3.9,1.4) -- (3.4,0.9) -- cycle;
        \fill [gray, opacity=0.5] (2.5,0.9) -- (3.4,0.9) -- (3.9,1.4) -- (3,1.4) -- cycle;

        \draw (2.5,0.5) -- (3.4,0.5) -- (3.9,1) -- (3.9,1.4) -- (3,1.4) -- (2.5,0.9) -- cycle;
        \draw (2.5,0.9) -- (3.4,0.9) -- (3.9,1.4);
        \draw (3.4,0.5) -- (3.4,0.9);

        \lp{}

        \fill [gray, opacity=0.5] (1.5,-0.5) rectangle (2.4,-0.1);
        \draw (1.5,-0.1) -- (2.4,-0.1) -- (2.4,-0.5);

        \fill [gray, opacity=0.5] (5,0.5) -- (5.5,1) -- (5.5,1.4) -- (5,0.9) -- cycle;
        \draw (5.5,1.4) -- (5,0.9) -- (5,0.5);

        \draw [->] (7,1) -- (8,1);
        
        \begin{scope}[shift={(8,0)}]

            \draw [red, thick] (2.2,0) -- (4.2,2) -- (4.2,4) -- (2.2,2) -- cycle;
            \draw [blue, thick] (0.8,0.8) -- (4.8,0.8) -- (4.8,2.8) -- (0.8,2.8) -- cycle;

            \draw [thick] (3.2,1) -- (3.2,3);
            \draw [thick] (2.8,0.8) -- (2.8,2.8);
            
            \fill [gray, opacity=0.5] (3.2,1.4) -- (2.7,0.9) -- (2.7,0.5) -- (3.2,1) -- cycle;
            \draw (3.2,1.4) -- (2.7,0.9) -- (2.7,0.5);
            \fill [gray, opacity=0.5] (2.8,1.2) rectangle (3.7,0.8);
            \draw (2.8,1.2) -- (3.7,1.2) -- (3.7,0.8);

            \draw (3,0.8) -- (3,2.8);

            \fill [green] (3,1) circle (3pt);
            \fill [green] (3,1.8) circle (3pt);

            \fill [blue] (4.4,1.2) circle (3pt);
            \fill [blue] (3.9, 2.5) circle(3pt);
        
            \fill [red] (3.4, 1.5) circle (3pt);
            \fill [red] (3.7, 3.2) circle (3pt);

        \end{scope}

    \end{tikzpicture}
\end{center}
            
\noindent On the left we choose a pair of planes through the code and on the right we restrict to these planes. The four circles in the blue/red plane (two blue/red and two green) are the support of an $X$/$Z$ operator associated with a codeword of the classical code, with the two green circles being the intersection of these operators. Notice that although they intersect on a qubit within the shaded region -- which in the HGP setting was sufficient to ensure they anticommute -- here they also intersect on another qubit vertically above this one and so the total intersection is even. 

Additionally, note that in general it is not even possible to find a basis for the classical codes that looks like the one shown above. This is because, while we can always find a basis such that $k$ bits somewhere in the code correspond to the $k$ encoded bits, arranging them in the fashion described above will in general require permutation of the vertices of the graph and this will not preserve the structure given to the graph by the lift (equivalence classes of vertices with respect to the covering map share the same column, etc). This is a problem because methods of ensuring commutativity of checks in the lifted product typically rely on this structure and if we start arbitrarily permuting vertices in the input we will usually not preserve commutativity in the output. 
\newline

\textbf{Code distance}

The methods used to lower bound the distance of HGP also do not work in this setting. Recall that the lower bound on the distance for HGP codes was derived using the fact that an operator with weight $< d_1$ must be supported on less than $d_1$ columns of $Q_1$, and so the punctured classical code containing only the bits corresponding to these columns must have codespace dimension $0$ (and similarly for rows of $Q_2$). Describing the classical base codes as $[n_i,k_i,d_i]$ and corresponding lifts as $[N_i,K_i,D_i]$ we could try to argue that in the lifted product an operator with weight $< D_1$ must be supported on less than $D_1$ planes through $Q_1$/less than $D_2^T$ planes through $Q_2$, which we might imagine like

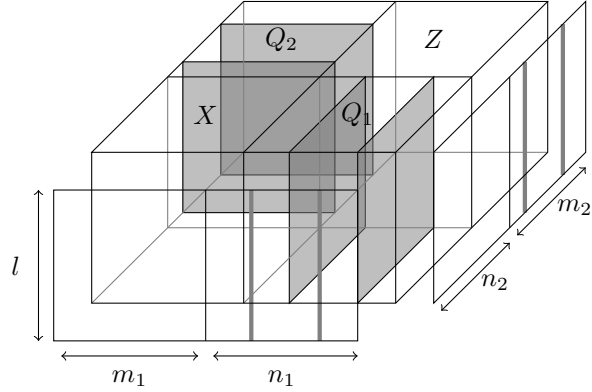
\begin{center}
    \begin{tikzpicture}
    
        \fill [gray, opacity=0.5] (1.7,1.7) rectangle (3.7,3.7);
        \draw (1.7,1.7) rectangle (3.7,3.7);
        \fill [gray, opacity=0.5] (1.2,1.2) rectangle (3.2,3.2);
        \draw (1.2,1.2) rectangle (3.2,3.2);

        \fill [gray, opacity=0.5] (2.6,0) -- (3.6,1) -- (3.6,3) -- (2.6,2) -- cycle;
        \draw (2.6,0) -- (3.6,1) -- (3.6,3) -- (2.6,2) -- cycle;
        \fill [gray, opacity=0.5] (3.5,0) -- (4.5,1) -- (4.5,3) -- (3.5,2) -- cycle;
        \draw (3.5,0) -- (4.5,1) -- (4.5,3) -- (3.5,2) -- cycle;

        \draw [gray,ultra thick] (2.1,-0.5) -- (2.1,1.5);
        \draw [gray,ultra thick] (3,-0.5) -- (3,1.5);

        \draw [gray,ultra thick] (6.2,1.7) -- (6.2,3.7);
        \draw [gray,ultra thick] (5.7,1.2) -- (5.7,3.2);

        \lp{}

    \end{tikzpicture}
\end{center}

\noindent and could then try think about the classical lifted codes punctured to only have support on the columns corresponding to these planes (shown as grey lines in the image), which is equivalent to a lift of a punctured base code. However, in general we can have $D_1 > n_1$ so this does not actually necessitate any puncturing at all. 

We could also try to argue that this operator must be supported on less than $D_1$ lines through $Q_1$/less than $D_2^T$ lines through $Q_2$, i.e. 

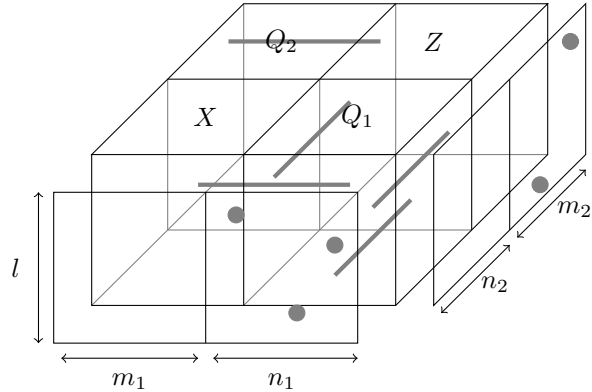
\begin{center}
    \begin{tikzpicture}

        \draw [gray, ultra thick] (1.4,1.6) -- (3.4,1.6);
        \draw [gray, ultra thick] (1.8,3.5) -- (3.8,3.5);
        
        \draw [gray, ultra thick] (2.4,1.7) -- (3.4,2.7);
        \draw [gray, ultra thick] (3.2,0.4) -- (4.2,1.4);
        \draw [gray, ultra thick] (3.7,1.3) -- (4.7,2.3);

        \lp{}

        \filldraw [gray] (1.9,1.2) circle (3pt);
        \filldraw [gray] (2.7,-0.1) circle (3pt);
        \filldraw [gray] (3.2,0.8) circle (3pt);

        \filldraw [gray] (5.9,1.6) circle (3pt);
        \filldraw [gray] (6.3,3.5) circle (3pt);

    \end{tikzpicture}
\end{center}

\noindent but puncturing only subsets of columns in the lifted classical codes will not generally preserve commutativity between checks of the product. Consider, for example

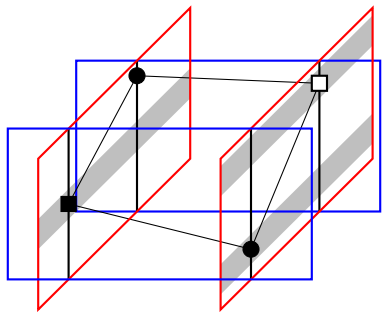
\begin{center}
    \begin{tikzpicture}
        
        \fill [gray, opacity=0.5] (3.2,0.2) -- (5.2,2.2) -- (5.2,2.6) -- (3.2,0.6) -- cycle;
        \fill [gray, opacity=0.5] (0.8,0.8) -- (2.8,2.8) -- (2.8,3.2) -- (0.8,1.2) -- cycle;
        \fill [gray, opacity=0.5] (3.2,1.5) -- (5.2,3.5) -- (5.2,3.9) -- (3.2,1.9) -- cycle;

        \draw [blue, thick] (1.3,1.3) -- (5.3,1.3) -- (5.3,3.3) -- (1.3,3.3) -- cycle;

        \draw (1.2,1.4) -- (3.6,0.8);
        \draw (4.5,3) -- (3.6,0.8);
        \draw (4.5,3) -- (2.1,3.1);
        \draw (2.1,3.1) -- (1.2,1.4);
        
        \draw [thick] (1.2,0.4) -- (1.2,2.4);
        \draw [thick] (2.1,1.3) -- (2.1,3.3);
        \draw [thick] (4.5,1.3) -- (4.5,3.3);

        \draw [red, thick] (0.8, 0) -- (2.8,2) -- (2.8,4) -- (0.8,2) -- cycle;

        \filldraw (1.1,1.3) rectangle (1.3,1.5);
        \filldraw (3.6,0.8) circle (3pt);
        \fill [white] (4.4,2.9) rectangle (4.6,3.1);    
        \draw [thick] (4.4,2.9) rectangle (4.6,3.1);    
        \filldraw (2.1,3.1) circle (3pt);
                
        \draw [thick] (3.6,0.4) -- (3.6,2.4);

        \draw [red, thick] (3.2,0) -- (5.2,2) -- (5.2,4) -- (3.2,2) -- cycle;
        \draw [blue, thick] (0.4,0.4) -- (4.4,0.4) -- (4.4,2.4) -- (0.4,2.4) -- cycle;
    
    \end{tikzpicture}
\end{center}

\noindent and so the product of classical codes punctured in this way cannot typically be used to infer properties of the code arising from the product of unpunctured classical codes. 
\newline

\textbf{Logical operations}

Due to the lack of a canonical basis for the logical operators in the general case the techniques for logic developed for HGP codes do not generally have a well-defined action on the logical qubits of LP codes. Additionally, techniques related to puncturing such as arxiv:2407.18490 will not be guaranteed to preserve distance when applied to these codes. Techniques based on folding, symmetries etc (arxiv:2202.06647, 2204.10812) will, as far as I can tell, usually still preserve the stabiliser group so there should be \textit{some} logical action, it is just not easy (or not possible?) to state what it will be in the general case. Given a specific code and known basis the action of these operations could be checked directly and you might still get a useful gate set. 
\newline

\textbf{Aysmptotic scaling}

Imagine that we somehow manage to find a family of asymptotically good classical input codes with parameters $[N_1,K_1 = O(N_1),D_1 = O(N_1)]$ and $[N_2,K_2 = O(N_2),D_2 = O(N_2)]$ (obtained from lifts of $[n_1,k_1,d_1]$ and $[n_2,k_2,d_2]$ codes) which yield a quantum code with the optimal parameters 
\begin{equation}
    [\![ n_1n_2l + m_1m_2l, O(n_1n_2l + m_1m_2l), \textrm{min}(D_1,D_2,D_1^T,D_2^T)]\!]
\end{equation}
To simplify the argument let's assume that $n_1 = n_2 = n$ and $m_1 = m_2 = m$, that we do not have $m \gg n$, and that $\textrm{min}(D_1,D_2,D_1^T,D_2^T) = D_1$ for any choice of $n$. We can then instead write these parameters as
\begin{equation}
    [\![ O(n^2l) , O(n^2l), O(nl)]\!]
\end{equation}
We can then see that if we grow the classical codes by growing the size of the base graphs while keeping the size of the lift constant then we have only a square-root scaling of the distance of the quantum code. On the other hand, if we grow the classical codes by increasing the lift size with a constant-size base graph then the quantum code is asymptotically good. In other words, families of good classical lifted codes are only useful in this (asymptotic) setting if the family is defined with a growing lift and fixed base graph, and not the reverse.

\subsection{Justification}
\input{lp_just}

\section{Balanced Product Codes}
Original paper arxiv:2012.09271

\subsection{Visualisation}
This one has perhaps the scariest reputation, but we have actually done most of the work necessary to understand it already. We just need a few more definitions first. 

Imagine we have two graphs, $A$ and $B$, and a group $H$ with a defined action on both of them. The action on $A$ must satisfy some conditions, namely 

\begin{itemize} 
    \item The action of $H$ on $A$ it must be \textit{free}, meaning that for any $a \in A$ and $h_1, h_2 \in H$, $h_1 \cdot a = h_2 \cdot a$ only if  $h_1 = h_2$.
    \item For any $a \in A$ and $h \in H$ there must be no edge between $a$ and $h \cdot a$. 
\end{itemize}

\noindent while the action on $B$ can be any valid action. For example, we might have that $A$ is the length-6 cycle graph, $B$ is the length-4 cycle graph and $H$ is the cyclic group $\mathbb{Z}_3 = \{0,z,z^2\}$, with action of its generating element $z$ defined on these graphs like

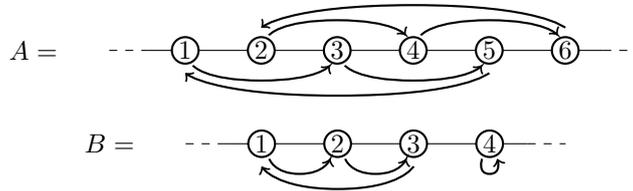
\begin{center}
    \begin{tikzpicture}
        \node at (0,0) {$A = $};
        
        \draw [dashed] (1,0) -- (1.5,0);
        \draw (1.5,0) -- (7.5,0);
        \draw [dashed] (7.5,0) -- (8,0);

        \begin{scope}[shift={(0,0)}]
            \fill [white] (2,0) circle (5pt);
            \draw [thick] (2,0) circle (5pt);
            \node at (2,0) {$1$};
        \end{scope}
        \begin{scope}[shift={(1,0)}]
            \fill [white] (2,0) circle (5pt);
            \draw [thick] (2,0) circle (5pt);
            \node at (2,0) {$2$};
        \end{scope}
        \begin{scope}[shift={(2,0)}]
            \fill [white] (2,0) circle (5pt);
            \draw [thick] (2,0) circle (5pt);
            \node at (2,0) {$3$};
        \end{scope}
        \begin{scope}[shift={(3,0)}]
            \fill [white] (2,0) circle (5pt);
            \draw [thick] (2,0) circle (5pt);
            \node at (2,0) {$4$};
        \end{scope}
        \begin{scope}[shift={(4,0)}]
            \fill [white] (2,0) circle (5pt);
            \draw [thick] (2,0) circle (5pt);
            \node at (2,0) {$5$};
        \end{scope}
        \begin{scope}[shift={(5,0)}]
            \fill [white] (2,0) circle (5pt);
            \draw [thick] (2,0) circle (5pt);
            \node at (2,0) {$6$};
        \end{scope}
        
        \draw [->, thick] plot [smooth, tension=1.5] coordinates {(2.1,-0.2) (3,-0.4) (3.9,-0.2)};
        \draw [->, thick] plot [smooth, tension=1.5] coordinates {(4.1,-0.2) (5,-0.4) (5.9,-0.2)};
        \draw [->, thick] plot [smooth, tension=1.5] coordinates {(6,-0.3) (4,-0.6) (2,-0.3)};

        \draw [->, thick] plot [smooth, tension=1.5] coordinates {(3.1,0.2) (4,0.4) (4.9,0.2)};
        \draw [->, thick] plot [smooth, tension=1.5] coordinates {(5.1,0.2) (6,0.4) (6.9,0.2)};
        \draw [->, thick] plot [smooth, tension=1.5] coordinates {(7,0.3) (5,0.6) (3,0.3)};

    \end{tikzpicture}
\end{center}

\begin{center}
    \begin{tikzpicture}
        \node at (0,0) {$B = $};
        
        \draw [dashed] (1,0) -- (1.5,0);
        \draw (1.5,0) -- (5.5,0);
        \draw [dashed] (5.5,0) -- (6,0);
        
        \begin{scope}[shift={(0,0)}]
            \fill [white] (2,0) circle (5pt);
            \draw [thick] (2,0) circle (5pt);
            \node at (2,0) {$1$};
        \end{scope}
        \begin{scope}[shift={(1,0)}]
            \fill [white] (2,0) circle (5pt);
            \draw [thick] (2,0) circle (5pt);
            \node at (2,0) {$2$};
        \end{scope}
        \begin{scope}[shift={(2,0)}]
            \fill [white] (2,0) circle (5pt);
            \draw [thick] (2,0) circle (5pt);
            \node at (2,0) {$3$};
        \end{scope}
        \begin{scope}[shift={(3,0)}]
            \fill [white] (2,0) circle (5pt);
            \draw [thick] (2,0) circle (5pt);
            \node at (2,0) {$4$};
        \end{scope}
        
        \draw [->, thick] plot [smooth, tension=1.5] coordinates {(2.1,-0.2) (2.5,-0.4) (2.9,-0.2)};
        \draw [->, thick] plot [smooth, tension=1.5] coordinates {(3.1,-0.2) (3.5,-0.4) (3.9,-0.2)};
        \draw [->, thick] plot [smooth, tension=1.5] coordinates {(4,-0.3) (3,-0.6) (2,-0.3)};

        \draw [->, thick] plot [smooth, tension=3] coordinates {(4.9,-0.2) (5,-0.4) (5.1,-0.2)};

    \end{tikzpicture}
\end{center}

\noindent where I have numbered the vertices for reference later. We will also need the notion of a quotient graph, for example $A/H$,

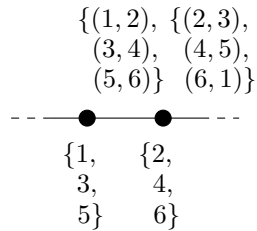
\begin{center}
    \begin{tikzpicture}
        
        \draw [dashed] (0,0) -- (0.5,0);
        \draw (0.5,0) -- (2.5,0);
        \draw [dashed] (2.5,0) -- (3,0);

        \filldraw (1,0) circle (3pt);
        \filldraw (2,0) circle (3pt);

        \node at (0.9,-0.5) {$\{1,$};
        \node at (1,-0.9) {$3,$};
        \node at (1.05,-1.3) {$5\}$};

        \node at (1.9,-0.5) {$\{2,$};
        \node at (2,-0.9) {$4,$};
        \node at (2.05,-1.3) {$6\}$};

        \node at (1.55,0.5) {$(5,6)\}$};
        \node at (1.5,0.9) {$(3,4),$};
        \node at (1.4,1.3) {$\{(1,2),$};

        \node at (2.75,0.5) {$(6,1)\}$};
        \node at (2.7,0.9) {$(4,5),$};
        \node at (2.6,1.3) {$\{(2,3),$};

    \end{tikzpicture}
\end{center}
                            
\noindent which has one vertex/edge for each equivalence class of vertices/edges in $A$ under the action of $H$ (vertex equivalence classes are below the graph and edge classes are above). Note that the action of the group on edges is inherited from the action on vertices as an edge can be viewed as a pair of vertices. 

To obtain the balanced product of these graphs we first arrange them in 2D as with lifted graphs in the previous section, so that vertices in the same equivalence class under the action of the group share the same column. Additionally we choose a ``basepoint'' for each equivalence class, which is a vertex to which we associate the identity element of the group. Every other vertex in the equivalence class then inherits its own associated element of the group, which is the element maps the basepoint to this vertex under the action of the group. Vertices in different equivalence classes with the same associated element of $H$ should lie in the same row in the 2D layout. For example, using the graphs above we have

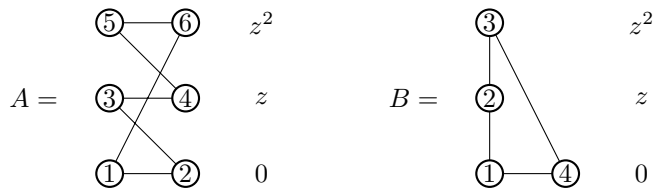
\begin{center}
    \begin{tikzpicture}

        \node at (1,1) {$A = $};
        
        \draw (2,0) -- (3,0) -- (2,1) -- (3,1) -- (2,2) -- (3,2) -- cycle;

        \begin{scope}[shift={(0,0)}]
            \fill [white] (2,0) circle (5pt);
            \draw [thick] (2,0) circle (5pt);
            \node at (2,0) {$1$};
        \end{scope}
        \begin{scope}[shift={(1,0)}]
            \fill [white] (2,0) circle (5pt);
            \draw [thick] (2,0) circle (5pt);
            \node at (2,0) {$2$};
        \end{scope}
        \begin{scope}[shift={(0,1)}]
            \fill [white] (2,0) circle (5pt);
            \draw [thick] (2,0) circle (5pt);
            \node at (2,0) {$3$};
        \end{scope}
        \begin{scope}[shift={(1,1)}]
            \fill [white] (2,0) circle (5pt);
            \draw [thick] (2,0) circle (5pt);
            \node at (2,0) {$4$};
        \end{scope}
        \begin{scope}[shift={(0,2)}]
            \fill [white] (2,0) circle (5pt);
            \draw [thick] (2,0) circle (5pt);
            \node at (2,0) {$5$};
        \end{scope}
        \begin{scope}[shift={(1,2)}]
            \fill [white] (2,0) circle (5pt);
            \draw [thick] (2,0) circle (5pt);
            \node at (2,0) {$6$};
        \end{scope}
        
        \node at (4,0) {$0$};
        \node at (4,1) {$z$};
        \node at (4,2) {$z^2$};

        \node at (6,1) {$B =$};

        \draw (7,0) -- (7,2) -- (8,0) -- cycle;

        \begin{scope}[shift={(5,0)}]
            \fill [white] (2,0) circle (5pt);
            \draw [thick] (2,0) circle (5pt);
            \node at (2,0) {$1$};
        \end{scope}
        \begin{scope}[shift={(5,1)}]
            \fill [white] (2,0) circle (5pt);
            \draw [thick] (2,0) circle (5pt);
            \node at (2,0) {$2$};
        \end{scope}
        \begin{scope}[shift={(5,2)}]
            \fill [white] (2,0) circle (5pt);
            \draw [thick] (2,0) circle (5pt);
            \node at (2,0) {$3$};
        \end{scope}
        \begin{scope}[shift={(6,0)}]
            \fill [white] (2,0) circle (5pt);
            \draw [thick] (2,0) circle (5pt);
            \node at (2,0) {$4$};
        \end{scope}
        
        \node at (9,0) {$0$};
        \node at (9,1) {$z$};
        \node at (9,2) {$z^2$};

    \end{tikzpicture}
\end{center}

\noindent where I have selected vertices $1$ and $2$ as basepoints for $A$ and $1$ and $4$ as basepoints for $B$. Now we can take a 3D product of these 2D graphs similar to what we did with the lifted product, which in this case would look like 

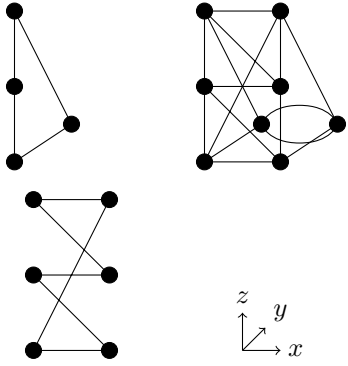
\begin{center}
    \begin{tikzpicture}
        
        \begin{scope}[shift={(-0.75,0.5)}] 
            \filldraw (0,0) circle (3pt);
            \filldraw (1,0) circle (3pt);
            \filldraw (0,1) circle (3pt);
            \filldraw (1,1) circle (3pt);
            \filldraw (0,2) circle (3pt);
            \filldraw (1,2) circle (3pt);

            \draw (0,0) -- (1,0) -- (0,1) -- (1,1) -- (0,2) -- (1,2) -- cycle;
        \end{scope}

        \filldraw (-1,3) circle (3pt);
        \filldraw (-1,4) circle (3pt);
        \filldraw (-1,5) circle (3pt);
        \filldraw (-0.25,3.5) circle (3pt);

        \draw (-1,3) -- (-1,5) -- (-0.25,3.5) -- cycle;

        \begin{scope}[shift={(1.5,3)}]
            \filldraw (0,0) circle (3pt);
            \filldraw (1,0) circle (3pt);
            \filldraw (0,1) circle (3pt);
            \filldraw (1,1) circle (3pt);
            \filldraw (0,2) circle (3pt);
            \filldraw (1,2) circle (3pt);

            \draw (0,0) -- (1,0) -- (0,1) -- (1,1) -- (0,2) -- (1,2) -- cycle;
        \end{scope}
        
        \filldraw (2.25,3.5) circle (3pt);
        \filldraw (3.25,3.5) circle (3pt);

        \draw (1.5,3) -- (1.5,5) -- (2.25,3.5) -- cycle;
        \draw (2.5,3) -- (2.5,5) -- (3.25,3.5) -- cycle;
        \draw plot [smooth, tension=1.5] coordinates {(2.25,3.5) (2.75,3.75) (3.25,3.5)};
        \draw plot [smooth, tension=1.5] coordinates {(2.25,3.5) (2.75,3.25) (3.25,3.5)};
    
        \begin{scope}[shift={(-2,0.5)}]
            \draw [->] (4,0) -- (4.5,0);
            \draw [->] (4,0) -- (4.3,0.3);
            \draw [->] (4,0) -- (4,0.5);
            \node at (4.7,0) {$x$};
            \node at (4.5,0.5) {$y$};
            \node at (4,0.7) {$z$};
        \end{scope}

    \end{tikzpicture}
\end{center}

Because the action of $H$ is free on $A$, every $yz$ plane of the product looks like a copy of $B$, exactly as with the lifted product. In cases where $H$ acts freely on the vertices of an equivalence class of $B$ (e.g. in the class containing vertices $1$, $2$ and $3$) then the corresponding $xz$ plane in the product looks like a copy of $A$, again, exactly as with the lifted product. The difference comes when we have equivalence classes of $B$ on which $H$ does not act freely, e,g, vertex $4$. In this case there are some elements of $H$ (in this case $z$) whose action on the basepoint of this class is trivial, and in the planes corresponding to these classes we will have a copy of $A$ quoiented by the subgroup generated by the elements of $H$ which act trivially on the basepoint (in this case the full group). 

To obtain a quantum code we just need to input two classical Tanner graphs to the product. There will be an additional requirement that bits must be mapped to bits and checks to checks by the action of $H$ on these graphs. As usual, the $X$ checks of the quantum code will be products of a check in the first graph and a bit in the second, $Z$ checks will be the reverse, and qubits will be either the product of two checks or two bits. In general we might view a balanced product code like this

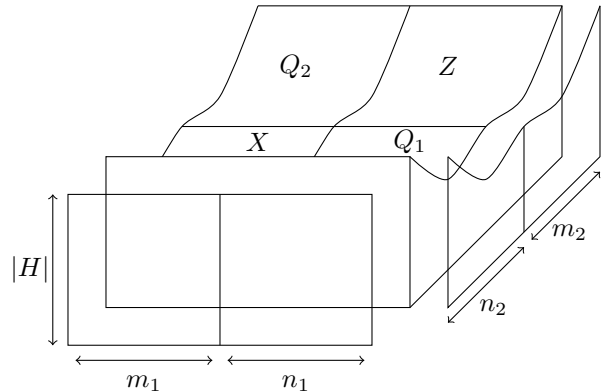
\begin{center}
    \begin{tikzpicture}
        \draw (4,0) -- (6,2) -- (6,4) -- (2,4);
        \draw (1,2.4) -- (5,2.4);
        
        \draw plot [smooth, tension=0.5] coordinates {(0,2) (0.5,1.7) (1,2.4) (1.5,2.8) (2,4)};
        \draw plot [smooth, tension=0.5] coordinates {(2,2) (2.5,1.7) (3,2.4) (3.5,2.8) (4,4)};
        \draw plot [smooth, tension=0.5] coordinates {(4,2) (4.5,1.7) (5,2.4) (5.5,2.8) (6,4)};
       
        \fill [white] (0,0) rectangle (4,2); 
        \draw (0,0) rectangle (4,2);

        \draw (-0.5,-0.5) rectangle (3.5,1.5);
        \draw (1.5,-0.5) -- (1.5,1.5);

        \draw (4.5,2) -- (4.5,0) -- (6.5,2) -- (6.5,4);
        \draw plot [smooth, tension=0.5] coordinates {(4.5,2) (5,1.7) (5.5,2.4) (6,2.8) (6.5,4)};
        \draw (5.5,1) -- (5.5,2.4);

        \draw [<->] (-0.4,-0.7) -- (1.4,-0.7);
        \draw [<->] (1.6,-0.7) -- (3.4,-0.7);
        \draw [<->] (-0.7,-0.5) -- (-0.7,1.5);

        \node at (0.5,-1) {$m_1$};
        \node at (2.5,-1) {$n_1$};
        \node at (-1,0.5) {$|H|$};

        \draw [<->] (4.5,-0.2) -- (5.5,0.8);
        \draw [<->] (5.6,0.9) -- (6.5,1.8);
        
        \node at (5.1,0) {$n_2$};
        \node at (6.1,1) {$m_2$};

        \node at (2,2.2) {$X$};
        \node at (4,2.2) {$Q_1$};
        \node at (2.5,3.2) {$Q_2$};
        \node at (4.5,3.2) {$Z$};

    \end{tikzpicture}
\end{center} 

\noindent where $n_1$ and $m_1$ ($n_2$ and $m_2$) are the number of equivalence classes of bits and checks in $A$ ($B$). From this perspective it looks like there is no general way to calculate the number of qubits and stabilisers in the balanced product code, but actually this is not the case. The balanced product can be viewed as the quotient of the Cartesian/hypergraph product of $A$ and $B$ by the action of $H$ (this is stated in the original balanced product paper and also used in the justification for the visualisation in the next subsection), and additionally the fact that $H$ acts freely on $A$ is sufficient to make it act freely on $A \times B$. The size of this code is then simply the size of the corresponding hypergraph product code divided by $|H|$. 
\newline

\textbf{Stabilisers commute}

Unlike the lifted product, the stabilisers of a balanced product code are guaranteed to commute. To see why we can once again use the perspective that the balanced product is the quotient of the Cartesian product by the action of $H$. Imagine that we have a qubit $q_1 \in Q_1$ of $A \times_H B$ that is in the support of both an $X$ stabiliser $x$ and $Z$ stabiliser $z$. These then correspond to equivalence classes $[q_1]$, $[x]$ and $[z]$, of qubits/stabilisers in $A \times B$ such that every qubit of $[q_1]$ is connected to an $X$ stabiliser in $[x]$ and $Z$ stabiliser in $[z]$. 

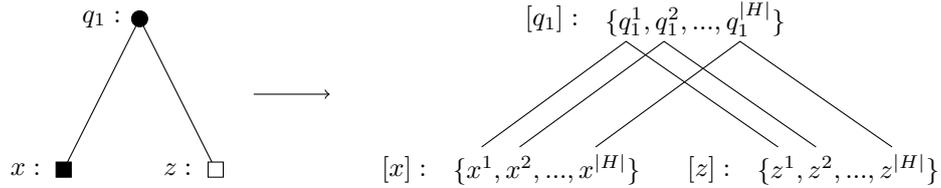
\begin{center}
    \begin{tikzpicture}
        
        \draw (0,0) -- (1,2) -- (2,0);
        \filldraw (-0.1,-0.1) rectangle (0.1,0.1);
        \node at (-0.5,0) {$x:$};
        \filldraw (1,2) circle (3pt);
        \node at (0.5,2) {$q_1:$};
        \fill [white] (1.9,-0.1) rectangle (2.1,0.1);
        \draw (1.9,-0.1) rectangle (2.1,0.1);
        \node at (1.5,0) {$z:$};

        \draw [->] (2.5,1) -- (3.5,1);

        \node [anchor = west] at (5,0) {$\{x^1,x^2,...,x^{|H|}\}$};
        \node [anchor = east] at (4.9,0) {$[x]:$};
        \node [anchor = west] at (7,2) {$\{q_1^1,q_1^2,...,q_1^{|H|}\}$};
        \node [anchor = east] at (6.9,2) {$[q_1]:$}; 
        \node [anchor = west] at (9,0) {$\{z^1,z^2,...,z^{|H|}\}$};
        \node [anchor = east] at (8.9,0) {$[z]:$};

        \draw (5.5,0.3) -- (7.4,1.7) -- (9.4,0.3);
        \draw (6,0.3) -- (7.9,1.7) -- (9.9,0.3);
        \draw (7,0.3) -- (8.9,1.7) -- (10.9,0.3);

    \end{tikzpicture}
\end{center}
        
For any triple $q_1^i$, $x^i$, $z^i$ there must then be a qubit $q_2^i \in Q_2$ connected to both $x^i$ and $z^i$ due to the structure of the HGP code, and because this vertex is unique for each triple (again due to the structure of the HGP code) all of these vertices belong to the same equivalence class. We can see this last fact by observing that $x^i$ and $z^i$ are connected to $q_2^i$ by edges $(x^i,q_2^i)$ and $(z^i,q_2^i)$ and acting on these with $h \in H$ we get $(h \cdot x^i,h \cdot q_2^i)$ and $(h \cdot z^i,h \cdot q_2^i)$. $h \cdot x^i$ and $h \cdot z^i$ are the $X$ and $Z$ checks of some other triple which share a unique vertex $h \cdot q_2^i \in Q_2$ which by definition is in the same equivalence class under $H$ as $q_2^i$. This gives us something like 

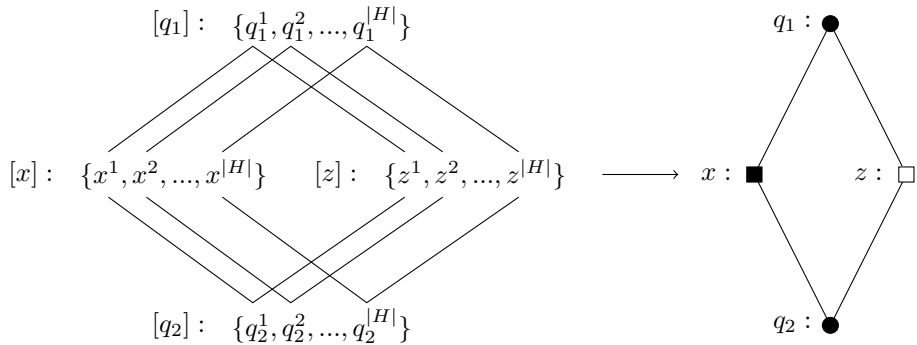
\begin{center}
    \begin{tikzpicture}
        
        \node [anchor = west] at (5,0) {$\{x^1,x^2,...,x^{|H|}\}$};
        \node [anchor = east] at (4.9,0) {$[x]:$};
        \node [anchor = west] at (7,2) {$\{q_1^1,q_1^2,...,q_1^{|H|}\}$};
        \node [anchor = east] at (6.9,2) {$[q_1]:$}; 
        \node [anchor = west] at (9,0) {$\{z^1,z^2,...,z^{|H|}\}$};
        \node [anchor = east] at (8.9,0) {$[z]:$};
        \node [anchor = west] at (7,-2) {$\{q_2^1,q_2^2,...,q_2^{|H|}\}$};
        \node [anchor = east] at (6.9,-2) {$[q_2]:$};

        \draw (5.5,0.3) -- (7.4,1.7) -- (9.4,0.3);
        \draw (6,0.3) -- (7.9,1.7) -- (9.9,0.3);
        \draw (7,0.3) -- (8.9,1.7) -- (10.9,0.3);
        \draw (5.5,-0.3) -- (7.4,-1.7) -- (9.4,-0.3);
        \draw (6,-0.3) -- (7.9,-1.7) -- (9.9,-0.3);
        \draw (7,-0.3) -- (8.9,-1.7) -- (10.9,-0.3);

        \draw [->] (12,0) -- (13,0);
    
        \begin{scope}[shift={(14,0)}]
            \draw (0,0) -- (1,2) -- (2,0) -- (1,-2) -- cycle;
            \filldraw (-0.1,-0.1) rectangle (0.1,0.1);
            \node at (-0.5,0) {$x:$};
            \filldraw (1,2) circle (3pt);
            \node at (0.5,2) {$q_1:$};
            \fill [white] (1.9,-0.1) rectangle (2.1,0.1);
            \draw (1.9,-0.1) rectangle (2.1,0.1);
            \node at (1.5,0) {$z:$};
            \filldraw (1,-2) circle (3pt);
            \node at (0.5,-2) {$q_2:$}; 
        \end{scope}

    \end{tikzpicture}
\end{center}

\noindent so there exists another vertex $q_2 \in Q_2$ in $A \times_H B$ that is connected to both $x$ and $z$. Since we can find such a unique $q_2$ for any $q_1$ connected to $x$ and $z$ the intersection of any $X$ and $Z$ stabiliser must always be even, ensuring that the stabilisers commute. 
\newline

\textbf{Encoded qubits, code distance, etc}

As far as I can tell, the techniques used in the HGP fail to generalise here in the same ways and for the same reasons as in the LP, so I won't go over it all again. But, speaking of the LP...
\newline

\textbf{Connection to the Lifted Product}

The balanced product coincides with the lifted product whenever $A$ can be viewed as a lift of $A/H$ and $B$ can be viewed as a lift of $B/H$, so operationally taking the quotient can be viewed as a covering map. In fact, this is always true for $A$ because the property that $H$ acts freely on $A$ combined with the property that $A$ contains no edges of the form $(a, h \cdot a)$ means that no two edges connected to the same vertex belong to the same equivalency class and so gives us the required bijection on neighbourhoods. We can therefore see that in cases where these same two properties are required of $B$ the lifted and balanced product will coincide. 

As far as I understand, neither of these operations is a strict generalisation of the other; the balanced product can function when one of the inputs is not a covering graph, while the lifted product cannot, and the lifted product can produce codes with non-commuting checks, while the balanced product cannot. Of course, this latter case is not particularly interesting since it does not define a valid quantum code. It is not obvious to me whether or not the set of valid lifted product codes is a subset of the set of balanced product codes, i.e. given any lifted product code with commuting stabilisers, can the covering maps on the input graphs always be viewed as quotients by a shared symmetry group. I leave this as an exercise for the reader, who doubtless has more mathematical aptitude than me.

\subsection{Justification}
This will go a little differently to the previous justification sections as there is not (to my knowledge) a general expression for the parity check matrices of a balanced product code. Instead I will use the fact that the balanced product is defined to be $A \times_H B = (A \times B)/H$ and argue that the graphs obtained from the definition of the product and the graphs I describe in the previous section really are the same. 

Lets first consider an example, which will be the same one from the previous section but considered in more detail, i.e. we have two graphs $A$ and $B$ acted on by the generating element $z$ of $\mathbb{Z}_3$ like

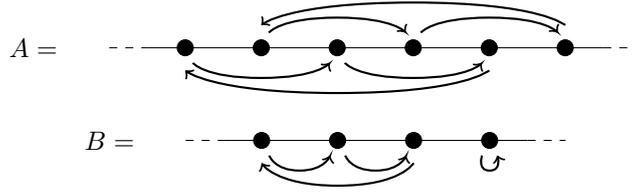
\begin{center}
    \begin{tikzpicture}
        \node at (0,0) {$A = $};
        
        \draw [dashed] (1,0) -- (1.5,0);
        \draw (1.5,0) -- (7.5,0);
        \draw [dashed] (7.5,0) -- (8,0);
        
        \filldraw (2,0) circle (3pt);
        \filldraw (3,0) circle (3pt);
        \filldraw (4,0) circle (3pt);
        \filldraw (5,0) circle (3pt);
        \filldraw (6,0) circle (3pt);
        \filldraw (7,0) circle (3pt);
                
        \draw [->, thick] plot [smooth, tension=1.5] coordinates {(2.1,-0.2) (3,-0.4) (3.9,-0.2)};
        \draw [->, thick] plot [smooth, tension=1.5] coordinates {(4.1,-0.2) (5,-0.4) (5.9,-0.2)};
        \draw [->, thick] plot [smooth, tension=1.5] coordinates {(6,-0.3) (4,-0.6) (2,-0.3)};

        \draw [->, thick] plot [smooth, tension=1.5] coordinates {(3.1,0.2) (4,0.4) (4.9,0.2)};
        \draw [->, thick] plot [smooth, tension=1.5] coordinates {(5.1,0.2) (6,0.4) (6.9,0.2)};
        \draw [->, thick] plot [smooth, tension=1.5] coordinates {(7,0.3) (5,0.6) (3,0.3)};

    \end{tikzpicture}
\end{center}

\begin{center}
    \begin{tikzpicture}
        \node at (0,0) {$B = $};
        
        \draw [dashed] (1,0) -- (1.5,0);
        \draw (1.5,0) -- (5.5,0);
        \draw [dashed] (5.5,0) -- (6,0);
        
        \filldraw (2,0) circle (3pt);
        \filldraw (3,0) circle (3pt);
        \filldraw (4,0) circle (3pt);
        \filldraw (5,0) circle (3pt);

        \draw [->, thick] plot [smooth, tension=1.5] coordinates {(2.1,-0.2) (2.5,-0.4) (2.9,-0.2)};
        \draw [->, thick] plot [smooth, tension=1.5] coordinates {(3.1,-0.2) (3.5,-0.4) (3.9,-0.2)};
        \draw [->, thick] plot [smooth, tension=1.5] coordinates {(4,-0.3) (3,-0.6) (2,-0.3)};

        \draw [->, thick] plot [smooth, tension=3] coordinates {(4.9,-0.2) (5,-0.4) (5.1,-0.2)};

    \end{tikzpicture}
\end{center}

\noindent and whose Cartesian and balanced products look like

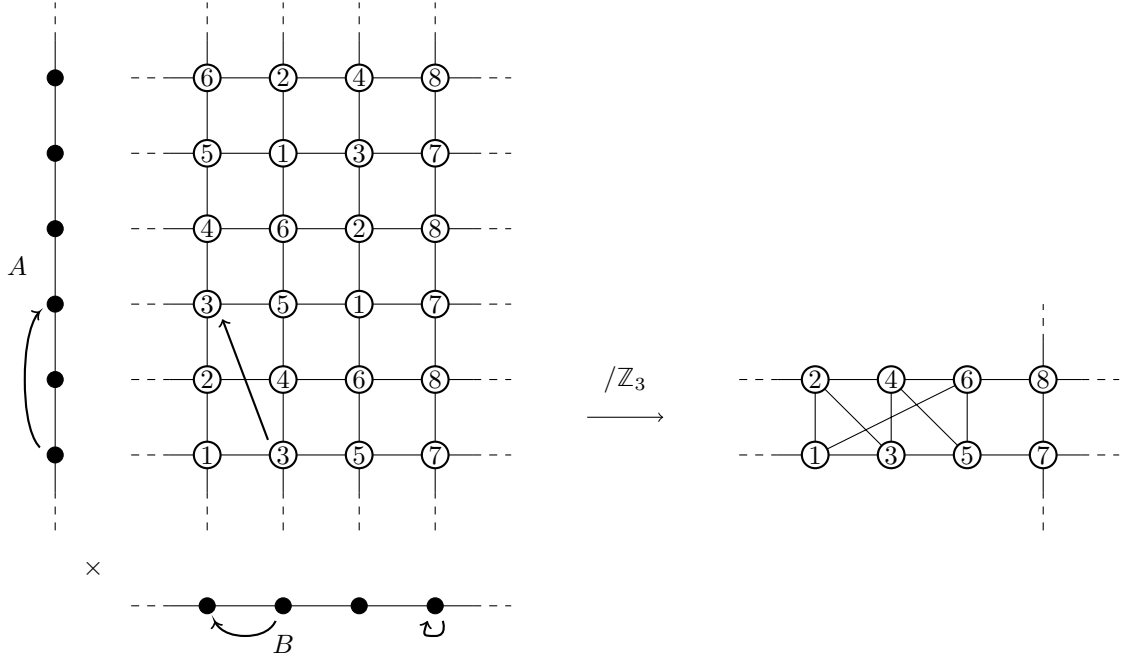
\begin{center}
    \begin{tikzpicture}

        \node at (-0.5,4.5) {$A$};
        \node at (3,-0.5) {$B$};

        \node at (0.5,0.5) {$\times$};
        
        \draw [dashed] (0,1) -- (0,1.5);
        \draw (0,1.5) -- (0,7.5);
        \draw [dashed] (0,7.5) -- (0,8);

        \filldraw (0,2) circle (3pt);
        \filldraw (0,3) circle (3pt);
        \filldraw (0,4) circle (3pt);
        \filldraw (0,5) circle (3pt);
        \filldraw (0,6) circle (3pt);
        \filldraw (0,7) circle (3pt);

        \draw [dashed] (1,0) -- (1.5,0);
        \draw (1.5,0) -- (5.5,0);
        \draw [dashed] (5.5,0) -- (6,0);

        \filldraw (2,0) circle (3pt);
        \filldraw (3,0) circle (3pt);
        \filldraw (4,0) circle (3pt);
        \filldraw (5,0) circle (3pt);

        \begin{scope}[shift={(0,2)}]
            \draw [dashed] (1,0) -- (1.5,0);
            \draw (1.5,0) -- (5.5,0);
            \draw [dashed] (5.5,0) -- (6,0);
        \end{scope}

        \begin{scope}[shift={(0,3)}]
            \draw [dashed] (1,0) -- (1.5,0);
            \draw (1.5,0) -- (5.5,0);
            \draw [dashed] (5.5,0) -- (6,0);
        \end{scope}

        \begin{scope}[shift={(0,4)}]
            \draw [dashed] (1,0) -- (1.5,0);
            \draw (1.5,0) -- (5.5,0);
            \draw [dashed] (5.5,0) -- (6,0);
        \end{scope}

        \begin{scope}[shift={(0,5)}]
            \draw [dashed] (1,0) -- (1.5,0);
            \draw (1.5,0) -- (5.5,0);
            \draw [dashed] (5.5,0) -- (6,0);
        \end{scope}
        
        \begin{scope}[shift={(0,6)}] 
            \draw [dashed] (1,0) -- (1.5,0);
            \draw (1.5,0) -- (5.5,0);
            \draw [dashed] (5.5,0) -- (6,0);
        \end{scope}
        
        \begin{scope}[shift={(0,7)}]
            \draw [dashed] (1,0) -- (1.5,0);
            \draw (1.5,0) -- (5.5,0);
            \draw [dashed] (5.5,0) -- (6,0);
        \end{scope}

        \begin{scope}[shift={(2,0)}]
            \draw [dashed] (0,1) -- (0,1.5);
            \draw (0,1.5) -- (0,7.5);
            \draw [dashed] (0,7.5) -- (0,8);
            
            \fill [white] (0,2) circle (5pt);
            \draw [thick] (0,2) circle (5pt);
            \node at (0,2) {$1$};
            \fill [white] (0,3) circle (5pt);
            \draw [thick] (0,3) circle (5pt);
            \node at (0,3) {$2$};
            \fill [white] (0,4) circle (5pt);
            \draw [thick] (0,4) circle (5pt);
            \node at (0,4) {$3$};
            \fill [white] (0,5) circle (5pt);
            \draw [thick] (0,5) circle (5pt);
            \node at (0,5) {$4$};
            \fill [white] (0,6) circle (5pt);
            \draw [thick] (0,6) circle (5pt);
            \node at (0,6) {$5$};
            \fill [white] (0,7) circle (5pt);
            \draw [thick] (0,7) circle (5pt);
            \node at (0,7) {$6$};
        \end{scope}

        \begin{scope}[shift={(3,0)}]
            \draw [dashed] (0,1) -- (0,1.5);
            \draw (0,1.5) -- (0,7.5);
            \draw [dashed] (0,7.5) -- (0,8);
            
            \fill [white] (0,2) circle (5pt);
            \draw [thick] (0,2) circle (5pt);
            \node at (0,2) {$3$};
            \fill [white] (0,3) circle (5pt);
            \draw [thick] (0,3) circle (5pt);
            \node at (0,3) {$4$};
            \fill [white] (0,4) circle (5pt);
            \draw [thick] (0,4) circle (5pt);
            \node at (0,4) {$5$};
            \fill [white] (0,5) circle (5pt);
            \draw [thick] (0,5) circle (5pt);
            \node at (0,5) {$6$};
            \fill [white] (0,6) circle (5pt);
            \draw [thick] (0,6) circle (5pt);
            \node at (0,6) {$1$};
            \fill [white] (0,7) circle (5pt);
            \draw [thick] (0,7) circle (5pt);
            \node at (0,7) {$2$};
        \end{scope}

        \begin{scope}[shift={(4,0)}]
            \draw [dashed] (0,1) -- (0,1.5);
            \draw (0,1.5) -- (0,7.5);
            \draw [dashed] (0,7.5) -- (0,8);
 
            \fill [white] (0,2) circle (5pt);
            \draw [thick] (0,2) circle (5pt);
            \node at (0,2) {$5$};
            \fill [white] (0,3) circle (5pt);
            \draw [thick] (0,3) circle (5pt);
            \node at (0,3) {$6$};
            \fill [white] (0,4) circle (5pt);
            \draw [thick] (0,4) circle (5pt);
            \node at (0,4) {$1$};
            \fill [white] (0,5) circle (5pt);
            \draw [thick] (0,5) circle (5pt);
            \node at (0,5) {$2$};
            \fill [white] (0,6) circle (5pt);
            \draw [thick] (0,6) circle (5pt);
            \node at (0,6) {$3$};
            \fill [white] (0,7) circle (5pt);
            \draw [thick] (0,7) circle (5pt);
            \node at (0,7) {$4$};
        \end{scope}
        
        \begin{scope}[shift={(5,0)}]
            \draw [dashed] (0,1) -- (0,1.5);
            \draw (0,1.5) -- (0,7.5);
            \draw [dashed] (0,7.5) -- (0,8);
         
            \fill [white] (0,2) circle (5pt);
            \draw [thick] (0,2) circle (5pt);
            \node at (0,2) {$7$};
            \fill [white] (0,3) circle (5pt);
            \draw [thick] (0,3) circle (5pt);
            \node at (0,3) {$8$};
            \fill [white] (0,4) circle (5pt);
            \draw [thick] (0,4) circle (5pt);
            \node at (0,4) {$7$};
            \fill [white] (0,5) circle (5pt);
            \draw [thick] (0,5) circle (5pt);
            \node at (0,5) {$8$};
            \fill [white] (0,6) circle (5pt);
            \draw [thick] (0,6) circle (5pt);
            \node at (0,6) {$7$};
            \fill [white] (0,7) circle (5pt);
            \draw [thick] (0,7) circle (5pt);
            \node at (0,7) {$8$};
        \end{scope}

        \draw [->, thick] plot [smooth, tension=1.5] coordinates {(2.9,-0.2) (2.5,-0.4) (2.1,-0.2)};

        \draw [->, thick] plot [smooth, tension=1.5] coordinates {(-0.2,2.1) (-0.4,3) (-0.2,3.9)};

        \draw [<-, thick] plot [smooth, tension=3] coordinates {(4.9,-0.2) (5,-0.4) (5.1,-0.2)};

        \draw [->, thick] (2.8,2.2) -- (2.2,3.8);

        \draw [->] (7,2.5) -- (8,2.5);
        \node at (7.5,3) {$/\mathbb{Z}_3$};

        \begin{scope}[shift={(8,2)}]
            \draw [dashed] (1,0) -- (1.5,0);
            \draw (1.5,0) -- (5.5,0);
            \draw [dashed] (5.5,0) -- (6,0);
        \end{scope}

        \begin{scope}[shift={(8,3)}]
            \draw [dashed] (1,0) -- (1.5,0);
            \draw (1.5,0) -- (5.5,0);
            \draw [dashed] (5.5,0) -- (6,0);
        \end{scope}
       
        \begin{scope}[shift={(1,0)}] 
            \draw (9,2) -- (9,3) -- (10,2) -- (10,3) -- (11,2) -- (11,3) -- cycle;
            \draw (12,1.5) -- (12,3.5);
            \draw [dashed] (12,1) -- (12,1.5);
            \draw [dashed] (12,3.5) -- (12,4);

            \fill [white] (9,2) circle (5pt);
            \draw [thick] (9,2) circle (5pt);
            \node at (9,2) {$1$};
            \fill [white] (9,3) circle (5pt);
            \draw [thick] (9,3) circle (5pt);
            \node at (9,3) {$2$};
            \fill [white] (10,2) circle (5pt);
            \draw [thick] (10,2) circle (5pt);
            \node at (10,2) {$3$};
            \fill [white] (10,3) circle (5pt);
            \draw [thick] (10,3) circle (5pt);
            \node at (10,3) {$4$};
            \fill [white] (11,2) circle (5pt);
            \draw [thick] (11,2) circle (5pt);
            \node at (11,2) {$5$};
            \fill [white] (11,3) circle (5pt);
            \draw [thick] (11,3) circle (5pt);
            \node at (11,3) {$6$};
            \fill [white] (12,2) circle (5pt);
            \draw [thick] (12,2) circle (5pt);
            \node at (12,2) {$7$};
            \fill [white] (12,3) circle (5pt);
            \draw [thick] (12,3) circle (5pt);
            \node at (12,3) {$8$};
        \end{scope}

    \end{tikzpicture}
\end{center}

\noindent where on the left we have the Cartesian product and on the right we have the quotient by $\mathbb{Z}_3$. Numbers show which vertices in the product belong to the same equivalence class. The action of $z$ on $B$ is reversed relative to the previous diagram because in general the action of $H$ on an element of $A \times B$ is defined to be $h \cdot (a,b) = (a \cdot h, h^{-1} \cdot b)$. We can see that this coincides with the visualisation we obtained in the previous section. Using this intuition we can continue to the general case. 

First consider the structure of the graph obtained in the visualisation. Each $yz$ plane of the visualisation is a copy of $B$ associated to an equivalence class $[a]$ of vertices in $A$, and so each vertex in the visualisation can be labelled by $([a],b)$. Within these planes there are edges between vertices $([a],b)$ and $([a],b')$ whenever there is an edge $(b,b')$ in $B$. Similarly, the $xz$ planes are copies of $A/S$ associated to equivalence classes $[b]$ of $B$, where $S$ is the subgroup of $H$ that acts trivially on the elements of $[b]$. We can therefore also give each vertex a label $(a_S,[b])$ and observe that there are edges between vertices $(a_S,[b])$ and $(a_S',[b])$ in \textit{these} planes whenever there is an edge $(a_S,a_S')$ in $A/S$. 

Now lets show that the balanced product has the same structure. When drawing the quotient graph of $(A \times B)/H$ we can choose one vertex $(a,b)$ of $A \times B$ from each equivalence class $[(a,b)]$ to act as a representative for the class. Because in general the action of $H$ on $A$ is free, vertices in the same row of the product will always belong to different equivalency classes\footnote{since these vertices have the forms $(a,b)$ and $(a,b')$ and so there can be no $h$ which maps between them, as $a = a \cdot h$ only when $h = 0$ which would mean $b' = b$} and so we can always choose $|A/H|$ of these rows, with each row corresponding to an equivalence class $[a]$, to obtain a full set of vertex representatives for $(A \times B)/H$. We can then label each of the vertices in this row $([a],b)$ and observe that there is an edge between $([a],b)$ and $([a],b')$ in $A \times_H B$ whenever there is an edge $(b,b')$ in $B$.

On the other hand, vertices in the same column do not necessarily all belong to different equivalence classes as the action of $H$ on $B$ is not required to be free. Specifically, if we have $h^{-1} \cdot b = b$ for some nontrivial $h^{-1}$ then $(a,b)$ and $(a \cdot h, b)$ belong to the same equivalence class and the same column, and more generally if we have some subgroup $S$ of $H$ such that $s \cdot b = b ~ \forall ~ s \in S$ then all vertices $(a \cdot s, b), ~ s \in S$ are in the same equivalence class. As a result, the vertices from this column that we can use as representatives are only the vertices $(a_S,b)$, where $a_S$ is a representative of an equivalence class in $A/S$. There is one such (partial) column of representatives for each equivalence class in $B$ and so we can equivalently label the vertices of $A \times_H B$ as $(a_S,[b])$. Finally, observe that in each column of $A \times_H B$ there will be an edge between two vertices $(a_S,[b])$ and $(a_S',[b])$ whenever there is an edge $(a_S,a_S')$ in $A/S$.

\section{Further generalisations}

We have seen that both the lifted and balanced product can be interpreted as a kind of 3D product of graphs in which pairs of vertices in the input only have a corresponding vertex in the output when they have a common $z$ coordinate, resulting in smaller output sizes than in the Cartesian/hypergraph product. Unlike the Cartesian product this is not guaranteed to produce codes with commuting stabilisers, and so we can think of both the lifted and balanced product as simply providing ways to identify subsets of input graphs that result in commuting checks. Whether either of these strategies is the optimal one, or whether there exist other useful ways of choosing inputs, seems like an interesting question. 

Geometrically, the saving relative to the HGP comes from the fact that the input graphs are contained in rectangles with dimensions $x \times z$ or $y \times z$, where the sides of length $z$ share a common axis, and so the product is contained in a cuboid with volume $xyz$ as opposed to a square of area $xyz^2$ as in the HGP. We might then ask if it is possible to achieve even greater savings by using an even higher dimensional product, e.g. we could imagine a 4D product whose inputs are a pair of graphs contained within cuboids $xzw$ and $yzw$ where the sides of area $zw$ lie in a common plane. The product would then have hypervolume $xyzw$ while the corresponding hypergraph product would have area $xyz^2w^2$. However, this is not a better saving that just stacking all planes with $w > 0$ on top of the $w=0$ plane so we instead get rectangles with areas $x(zw)$ and $y(zw)$ and product with volume $xy(zw)$, so it seems the 3D product is optimal from this perspective.  

\end{document}